\documentclass[11pt,showpacs,preprintnumbers,amsmath,amssymb]{revtex4}
\usepackage{epsfig,amsmath}
\usepackage{graphicx}
\usepackage{dcolumn}

\newcommand{\beq}{\begin{equation}} \newcommand{\eeq}{\end{equation}}
\newcommand{\beqa}{\begin{eqnarray}}
\newcommand{\eeqa}{\end{eqnarray}} 
 \newcommand{\rh}{\rho}

\newcommand{\non}{\nonumber}

 \def\prl#1{{ Phys.\ Rev.\
Lett.} {\bf#1}}

\begin{document}

\title{Exact Master Equation and Quantum Decoherence  of Two Coupled Harmonic
Oscillators in
a General Environment}

\author{Chung-Hsien Chou$^{1,2}$\footnote{Email Address:
chouch@phys.sinica.edu.tw}, Ting Yu$^{3}$\footnote{Email address:
ting@pas.rochester.edu} and B.~L. Hu$^{2}$\footnote{Email address:
blhu@umd.edu}} \affiliation{$^1$Center for Gravitation, Cosmology and
Quantum Physics, Institute of Physics, Academia Sinica,
Nankang, Taipei 11529, Taiwan\\
$^2$Joint Quantum Institute and Maryland Center for Fundamental
Physics,\\ Department of Physics, University of Maryland, College
Park, Maryland 20742-4111\\
$^3$Department of Physics and Astronomy, University of Rochester,
Rochester, New York 14627-0171}
\date{Second Version Dec. 2, 2007}

\begin{abstract}
In this paper we derive an exact master equation for two coupled
quantum harmonic oscillators interacting via bilinear coupling with a
common environment at arbitrary temperature made up of many harmonic
oscillators with a general spectral density function. We first show a
simple derivation based on the observation that the two-harmonic
oscillator model can be effectively mapped into that of a single
harmonic oscillator in a general environment plus a free harmonic
oscillator. Since the exact one harmonic oscillator master equation
is available [Hu, Paz and Zhang, Phys. Rev. D \textbf{45}, 2843
(1992)], the exact master equation with all its coefficients for this
two harmonic oscillator model can be easily deduced from the known
results of the single harmonic oscillator case.  In the second part
we give an influence functional treatment of  this model and provide
explicit expressions for the evolutionary operator of the reduced
density matrix which are useful for the study of decoherence and
disentanglement issues. We show three applications of this master
equation: on the decoherence and disentanglement of two harmonic
oscillators due to their interaction with a common environment under
Markovian approximation, and a derivation of the uncertainty
principle at finite temperature for a composite object, modeled by
two interacting harmonic oscillators.  The exact master equation for
two, and its generalization to $N$,  harmonic oscillators interacting
with a general environment are expected to be useful for the analysis
of quantum coherence, entanglement, fluctuations and dissipation of
mesoscopic objects towards the construction of a theoretical
framework for macroscopic quantum phenomena.
\end{abstract}

\maketitle
\newpage
\section{Introduction}
Macroscopic quantum coherence phenomena (MQP) manifested in double
slit experiments, micromechanical resonators, Bose-Einstein
condensates, Josephson junction circuits, mesoscopic systems, or
even mirrors (see, e.g., \cite{Arndt,BJK,Friedman,Armour,
Eisert,Mancini02,Mancini03,Marshall,Adler,Pinard,KRCSV,FGV,Bose,Blencowe,Burnettetaal2007})
is a subject of both  basic theoretical and practical application
interest. Theoretically it focuses on issues at the intersection
of two trunk lines of important inquires in physics: the relation
between the microscopic and the macroscopic world on the one hand,
and the relation between the quantum and the classical on the
other. Rapid recent advances in precision measurements with high
degree of control and adaptability in atomic-optical,
electro-mechanical, opto-mechanical, nano-material, magnetic-spin
and low temperature systems have provided the rationale and
substance for such theoretical investigations, and in some
emergent areas where high goals are set, such as the quest for
quantum information processing, even with some sense of urgency.

The issues of interest in MQP include quantum dissipation,
entanglement, teleportation, decoherence, noise, correlation and
fluctuations. A familiar model which one could use to address many of
these issues is the quantum Brownian motion (QBM)
\cite{QBM,CalLeg,QBM1,HPZ,HalliwellYu96} and its dynamics described
by the master equation or the associated Langevin or Fokker-Planck
equations.  But since the systems of interest to MQP necessarily
involve many microscopic or mesoscopic constituents, a many-body
generalization of QBM is needed. In addition, since most of these
systems involve non-negligible correlations amongst their components,
quantum memory (non-Markovian) effects cannot be ignored. Even for
the well-studied single harmonic oscillator (1HO) QBM, Markovian
approximation is valid only for a high temperature Ohmic bath
\cite{CalLeg}. Fortunately an exact master (HPZ) equation \cite{HPZ}
for the 1HO with bilinear coupling to a general environment has been
found via several techniques ranging from the influence functional
\cite{HPZ} and Wigner function \cite{HalliwellYu96} to quantum
trajectories \cite{StrunzYu2004}.  The 1HO master equation for the
QBM is complex enough to encompass non-Markovian dynamics yet simple
enough to yield exact solutions. (See, e.g., \cite{FHR} and
references therein.)  The new challenge is to find the master
equation for $N$ oscillators in a general environment good for the
analysis of these issues in mesoscopic physics.

In this paper we show the derivation of such an equation for two
coupled harmonic oscillators (2HO). A key observation is that this
problem can be mapped into that of a single harmonic oscillator in a
general environment plus a free harmonic oscillator. Since the master
equation with all its coefficients for the 1HO QBM is known
\cite{HPZ,HalliwellYu96}  one can derive the master equation for the
2HO QBM easily from them. As an application of this model, we can
deduce the decoherence properties of the 2HO system following the
similar pattern of the 1HO.  As another example, we show explicitly
how, in some parameter choice, under the Markovian limit, an
entangled state evolves into a separable state in a finite time.

The results derived in this paper may be deduced by intuitive
reasoning, but we are not aware of any theoretical study which
yields our results. Our aim here is to provide a proof, or at
least a plausibility argument, to the effect that the center of
mass coordinate is the one most sensitive to the environmental
influence. This model and its generalization to $N$ harmonic
oscillators are expected to be useful for the analysis of quantum
coherence, entanglement, fluctuations and dissipation of
mesoscopic and macroscopic objects.

The paper is organized as follows: in Section~\ref{modelequation}, we
consider the dynamics of two harmonic oscillators coupled to a common
heat bath. By employing the center of mass and relative coordinates
we show how to derive the master equations of the two coupled
Brownian particles.  In Section~\ref{propagatorsection} we use the
influence functional method and derive an exact form of the
propagators for the reduced density matrices. These results are
expected to be useful for analyzing general statistical mechanical
properties of quantum open systems. In Section~\ref{application} we
give three examples as applications of this master equation: the
quantum decoherence and disentanglement of two interacting Brownian
oscillators in a general environment, and the uncertainty relation at
finite temperature for a composite object modeled by two interacting
oscillators. In Section~\ref{summarysection} we mention a few more
problems and physical issues where the results from this work can be
usefully applied to for their analysis and further extension of the
present study. Technical details are relegated to the two appendices.

\section{The Model and the Exact Master Equation}
\label{modelequation}

Quantum Brownian motion (QBM) of a damped harmonic oscillator
bilinearly coupled to a bath of harmonic oscillators has been
studied for decades, notably by Feynman-Vernon and Caldera-Leggett
using path integral techniques \cite{QBM,CalLeg}. For such a model
 an exact master equation can be deduced without making the Markovian
approximation \cite{HPZ}. The purpose of this section is to extend
the well-known Brownian motion model into the case where the
system of interest contains two coupled harmonic oscillators.

\subsection{The Model} \label{model}
The Hamiltonian of the total system consisting of a system (sys)
of two mutually coupled harmonic oscillators of equal mass $M$ and
frequency $\Omega$ interacting with a bath (bath) of $N_B$
harmonic oscillators of masses $m_n$ and frequencies $\omega_n$ in
an equilibrium state at a finite temperature $T$ can be formally
written as, \beq \label{tot}
 H_{\rm tot} = H_{\rm sys} + H_{\rm bath} + H_{\rm int},
 \eeq
where
\beq
 H_{\rm sys} = \frac{P_1^2}{2M} + \frac{1}{2}M \Omega^2 x_1^2 + \frac{P_2^2}{2M} +
 \frac{1}{2}M \Omega^2 x_2^2 + \kappa (x_1-x_2)^k
\eeq is the system Hamiltonian for the two system oscillators of
interest, with $(x_1, x_2)$ displacements, conjugate momenta
$(P_1, P_2)$ and coupling constant $\kappa$,
\begin{eqnarray}
 H_{\rm bath} = \sum_{n=1}^{N_B} ( \frac{p_n^2}{2m_n} + \frac{1}{2}m_n \omega_n^2
q_n^2 )
\end{eqnarray} is  the  bath Hamiltonian with displacement $q_n$
for the $n^{th}$ oscillator and conjugate momentum $p_n$ and \beq
 H_{\rm int } =( x_1 + x_2  )\sum_{n=1}^{N_B} C_n q_n
\eeq is  the interaction Hamiltonian between the system and the
bath. Here for simplicity, we have  assumed that the two harmonic
oscillators are coupled with the same coupling constants $C_n$ to
the bath oscillators.

Our primary focus in this paper is to derive an exact master
equation for the two coupled harmonic oscillators. Since the two
harmonic oscillators interact with a common thermal bath, there
will be induced coupling between the two harmonic oscillators even
when initially they are uncoupled. Thus, the master equation for
2HO QBM is not simply the addition of the two master equations for
1HO QBM. It must account for the mutual interactions between the
two Brownian particles introduced by their coupling to the common
heat bath. Of interest is a comparison with the model that
consists of 2HO each in its own heat bath. In our model, the
coupling to a common heat bath can give rise to several new
features, of particular interest here is the generation of
entanglement between the two Brownian particles due to the
back-action of the heat bath on the system
\cite{Kim2002,Dan,Kim,Ficek}.

However, as is well-known for classical mechanics, the dynamics of
an N body quantum open system can be made simpler by changing the
N body coordinates to that of their center of mass (cm) and
relative (rel) coordinates. Here, the difference is that the N
harmonic oscillators (NHO) are coupled with an environment and we
seek a quantum mechanical treatment. A quantum mechanical theory
of N body dynamics forms the theoretical basis for treating MQP.
In this paper we treat the 2HO case. We will show in what follows
that the exact master equation for the two coupled harmonic
oscillators can be obtained directly from the master equation for
the single harmonic oscillator, known as the Hu-Paz-Zhang (HPZ)
master equation.

Let us first rewrite the total Hamiltonian in terms of a set of new
variables $X, x, P, p$ defined as
\begin{eqnarray}
\label{trans1}
 X = \frac{1}{2} ( x_1 + x_2 ) , \quad x =  x_1 - x_2 ,
\end{eqnarray}
\begin{eqnarray}
\label{trans2}
 P =   P_1 + P_2, \quad p =  \frac{1}{2}( P_1 - P_2 ),
\end{eqnarray} and the new masses $M_1=2M, M_2=M/2.$
In terms of these new variables the Hamiltonian (\ref{tot}) takes the
following form:
\begin{equation}
H_{\rm sys}  =  H_{\rm cm}+H_{\rm rel}
\end{equation}
where
\begin{equation}
H_{\rm cm} =   \frac{P^2}{2M_1} + \frac{1}{2}M_1 \Omega^2 X^2,
\end{equation}
\begin{equation}
H_{\rm rel}= \frac{p^2}{2M_2} + \frac{1}{2}M_2 \Omega^2
 x^2 +\kappa x^k,
\label{H1H2}
\end{equation}
and
\begin{eqnarray}
 H_{\rm int} &=&( x_1 + x_2  )\sum_{n=1}^{N_B} C_n
 q_n =  2 X \sum_{n=1}^{N_B} C_n q_n = X \sum_{n=1}^{N_B} \tilde{C_n} q_n
\end{eqnarray}
where $\tilde{C_n} = 2 C_n$ are modified coupling constants. Since
(\ref{trans1}) and (\ref{trans2}) are canonical transformations, all
the commutators are preserved, and it is easy to check that
\begin{eqnarray}
 [ X, P ] =  [x, p ]=i\hbar,\,\,\,  [ P, x ] =  [ p, X ] =[X,x]=[P,p]=
 0.
\end{eqnarray}
We see that the fictitious particle with mass $M_2$ and dynamical
variables  $x, p$ has no interaction with either the cm particle
with mass $M_1$ with canonical variables $X,P$ or the oscillators
of the heat bath with canonical variables $q_n$.

The total Hamiltonian $H_{\rm tot}$ in (1) can now be written as $H_{\rm
tot} = H'_{\rm tot} + H_{\rm rel}$ with a new effective total
Hamiltonian  \beqa \label{tot1}
H_{\rm tot}' &=& H_{\rm cm} +H_{\rm int} +H_{\rm bath}\non \\
&&= \frac{P^2}{2M_1} + \frac{1}{2}M_1 \Omega^2 X^2+ X
\sum_{n=1}^{N_B} \tilde{C_n} q_n+\sum_{n=1}^{N_B} (
\frac{p_n^2}{2m_n} + \frac{1}{2}m_n \Omega_n^2 q_n^2 ). \eeqa This
Hamiltonian is formally the  same as the Hamiltonian for the
single harmonic oscillator in cm variables $(X,P)$ coupled to the
heat bath with coupling constants $\tilde C_n$.  Note that for
this case the spectral density ${\tilde I}(\omega)$ is given by:
\begin{eqnarray}
{\tilde I}(\omega)  = \pi  \sum_{n=1}^{N_B} \frac{\tilde{
C_n}^2}{2m_n \omega_n} \delta(\omega - \omega_n),
\end{eqnarray} which differs from the original spectral density $I(\omega)$
by a numerical factor $4$.

\subsection{Density Matrix}

We now consider the dynamics of two coupled harmonic oscillators
interacting with a common heat bath. The density matrix $\rho$
evolves in time under the unitary operator:
\begin{eqnarray}
\label{evolution1}
 \rho(t) = \exp\left[{-i\frac{H_{\rm tot}t}{\hbar}} \right]\rho(0)
\exp\left[{i\frac{H_{\rm
 tot}t}{\hbar}}\right].
\end{eqnarray}
From (\ref{tot1}), it is easy to see that this evolution can be
decomposed into two parts, a dissipative evolution of the center
of mass system,
\begin{eqnarray}
\label{evolution2}
 \tilde \rho(t) = \exp\left[{-i\frac{H'_{\rm tot}t}{\hbar}}\right] \rho(0)
\exp\left[{i\frac{H'_{\rm
 tot}t}{\hbar}}\right],
\end{eqnarray}
and the unitary evolution of the free harmonic oscillator with mass
$M_1$ , \beq \label{evolution3}
 \rho(t) = \exp\left[{-i\frac{H_{\rm rel} t}{\hbar}}\right] \tilde \rho(t)
 \exp{\left[i\frac{H_{\rm rel}t}{\hbar}\right]},
\eeq where $H_{\rm rel}$ is the Hamiltonian for the 1HO system with
reduced mass $M_2=M/2$ and $x,p$ variables: \beq \label{H2}
 H_{\rm rel}=\frac{p^2}{2M_2} + \frac{1}{2}M_2 \Omega^2
 x^2 +\kappa x^k.
\eeq

For technical simplicity we make the usual assumption that the
initial state of the total system is uncorrelated,
\begin{eqnarray}
\rho(0) = \rho_{\rm sys}(0) \times \rho_{\rm bath}(0),
\end{eqnarray}
and that the heat bath is in a thermal equilibrium state at
temperature $T$.

\subsection{Exact Master Equation}

If we are interested in the detailed dynamics of the system but only
the coarse-grained effect of the bath we can work with the reduced
density matrix obtained by tracing $\rho$, the density matrix of the
total system described by (\ref{tot}),  over the bath variables
\cite{Kubo,Gardiner}: \beq \rh_r={\rm Tr}_{\rm bath}\rh(t). \eeq The
reduced density operator for the center of mass system is obtained in
a similar way, \beq \tilde\rh_r={\rm Tr}_{\rm bath}\tilde\rh(t).
 \eeq
where ${\tilde \rho}$  defined in (\ref{evolution2}) is the density
operator for the effective total system (\ref{tot1}). The
relationship between $\tilde\rho_r$ and $\rho_r$ is given by
\begin{eqnarray}
\label{relation}
 \rho_r(t) = \exp\left[{-i\frac{H_{\rm rel} t}{\hbar}}\right]{\tilde\rho}_r(t)
\exp\left[{i\frac{H_{\rm rel} t}{\hbar}}\right].
\end{eqnarray}

Tracing over the heat bath variables in (\ref{evolution2}) leads us
to  a HPZ type master equation for the center of mass variables
$X,P$:
\begin{equation}\label{masterosci}
\dot{\tilde\rho}_r = \frac{1}{i\hbar}[H_{\rm cm},\tilde\rho_r]
+\frac{a(t)}{2i\hbar}[X^2,\tilde\rho_r] +
\frac{b(t)}{2i\hbar}[X,\{P,\tilde\rho_r\}]
+\frac{c(t)}{\hbar^2}[X,[P,\tilde\rho_r]] -
\frac{d(t)}{\hbar^2}[X,[X,\tilde\rho_r]].
\end{equation}
Note here that $H_{\rm cm}$ defined in (\ref{H1H2}) is the
Hamiltonian for the center of mass variables $X,P$ only. This is the
exact master equation for $X,P$ interacting with a thermal heat bath
with the spectral density ${\tilde I}(\omega)$ rather than
$I(\omega)$.  As a consequence, the coefficients $a, b, c, d$ in the
above master equation satisfy the same types of equations given by
\cite{HPZ} (or \cite{HalliwellYu96}), only the coupling constants and
mass are different here.

From the  evolution equation (\ref{evolution3}), the required
master equation for the reduced density matrix $\rho_r(t)$ is thus
obtained,
\begin{equation}\label{masterosci1}
\dot\rho_r = \frac{1}{i\hbar}[H_{\rm sys},\rho_r]
+\frac{a(t)}{2i\hbar}[X^2,\rho_r] +
\frac{b(t)}{2i\hbar}[X,\{P,\rho_r\}]
+\frac{c(t)}{\hbar^2}[X,[P,\rho_r]] -
\frac{d(t)}{\hbar^2}[X,[X,\rho_r]].
\end{equation}
The only difference between Eq. (\ref{masterosci1}) and Eq.
(\ref{masterosci}) is that the unitary evolution is modified by the
fictitious harmonic oscillator $x,p$.

In terms of the original variables  $x_1,x_2, P_1, P_2$, we get
\beqa \dot\rho_r &=& \frac{1}{i\hbar}[H_{\rm sys},\rho_r]
+\frac{a(t)}{8i\hbar}[(x_1+x_2)^2,\rho_r] +
\frac{b(t)}{4i\hbar}[x_1+x_2,\{P_1+P_2,\rho_r\}]\nonumber \\
&&+\frac{c(t)}{2\hbar^2}[x_1+x_2,[P_1+P_2,\rho_r]] -
\frac{d(t)}{4\hbar^2}[x_1+x_2, [x_1+x_2,\rho_r]]. \eeqa This is the
exact master equation for the two coupled harmonic oscillators. In
the coordinate representation, \beq \rh_r(x_1,x_2,y_1,y_2)\equiv
\langle x_1,x_2|\rho_r|y_1,y_2\rangle, \eeq the master equation can
be easily written as:
\begin{eqnarray}
\label{mastercoord}
 i\hbar \frac{\partial \rho_r}{\partial t}  &=&
-\frac{\hbar^2}{2M}\left(\frac{\partial^2}{\partial x_1^2}  -
\frac{\partial^2}{\partial y_1^2} + \frac{\partial^2}{\partial
x_2^2} - \frac{\partial^2}{\partial y_2^2} \right) \rho_r +
\frac{1}{2}M\Omega^2
(x_1^2 - y_1^2 + x_2^2 - y_2^2 ) \rho_r \nonumber \\
&& + \frac{1}{2}M\delta \Omega^2(t)(x_1 - y_1 + x_2 -
y_2)\frac{1}{2}(x_1 +y_1 + x_2 +
y_2)\rho_r  \nonumber \\
&& -i \hbar \Gamma(t)(x_1 - y_1 + x_2 - y_2)\frac{1}{2}\left(
\frac{\partial}{\partial x_1} - \frac{\partial}{\partial y_1} +
\frac{\partial}{\partial x_2} -
\frac{\partial}{\partial y_2} \right)\rho_r  \nonumber \\
&& -i M \Sigma(t)(x_1 - y_1 + x_2 - y_2)^2 \rho_r \nonumber  \\
&& + \hbar \Delta(t)(x_1 - y_1 + x_2 - y_2)\left(
\frac{\partial}{\partial x_1} + \frac{\partial}{\partial y_1} +
\frac{\partial}{\partial x_2} + \frac{\partial}{\partial y_2}
\right)\rho_r.
\end{eqnarray}
A set of new notations in (\ref{mastercoord}) is introduced to
facilitate easy adoption of results from \cite{HPZ}. In particular,
 \beqa
a(t)&=&M\delta\Omega^2(t),\,\,\,\,
b(t) = 2 \Gamma(t),\\
c(t)&=&\Delta(t),\,\,\,\,
d(t)=\Sigma(t). \eeqa

It is often useful to use the Wigner function defined in phase space,
which is related to the reduced density matrix $\rho_r$ in the
following way: \beqa \label{wignerdensity} \tilde
W(x_1,x_2,P_1,P_2,t)&=&{1\over {(2\pi)^2}}\int du_1du_2\ e^{{i(u_1
P_1+u_2P_2)
/\hbar}}\nonumber\\
&&\times \rho_r\left(x_1-{u_1\over 2},x_2-{u_2\over 2};x_1+{u_1\over
2},x_2+{u_2\over 2},t\right). \eeqa In correspondence with
(\ref{mastercoord}) the Wigner function satisfies a Fokker-Planck
equation:
\begin{eqnarray}
\label{fp}
{\partial\tilde W \over\partial t}= & -&\sum_{i=1,2}\left({P_i\over
M}{\partial\tilde W\over \partial x_i} - M\Omega^2 x_i{\partial\tilde W \over
\partial P_i}\right)\nonumber \\
&+&M\delta\Omega^2(t)(x_1+x_2)\left(\frac{\partial}{\partial
P_1}+\frac{\partial}{\partial P_2}\right)\tilde W
                                    +2 \Gamma(t)\left(\frac{\partial}{\partial
P_1}+\frac{\partial}{\partial P_2}\right)[(P_1
+P_2)\tilde W] \nonumber \\
&+&\Sigma(t)\left(\frac{\partial}{\partial
P_1}+\frac{\partial}{\partial P_2}\right)^2\tilde W +
\Delta(t)\left(\frac{\partial}{\partial
P_1}+\frac{\partial}{\partial
P_2}\right)\left(\frac{\partial}{\partial
x_1}+\frac{\partial}{\partial x_2}\right)\tilde W.
\end{eqnarray}
The time-dependent functions $\delta\Omega^2(t), \Gamma(t),
\Delta(t), \Sigma(t)$ are derived following the same method used by
HPZ which can be found in Appendix A.5.

In deriving the exact master equation we assumed that the initial
state for the two harmonic oscillators is  a product of a function of
the relative coordinates and a function of the center of mass
coordinates. However, it can be easily shown that the derivation is
valid for an arbitrary initial state of the system regardless of the
condition of separability.

\subsection{Markov Approximations}

The derived master equation (\ref{mastercoord}) is exact, so it is
valid in both the Markovian and the non-Markovian regimes. Memory
effects due to the environment is encoded in the time-dependent
coefficients. In the high temperature ohmic bath limit, the
coefficients become constants and the spectral density has the form:
\beq {I}(\omega)=M_1\gamma
\omega\exp\left(-\frac{\omega^2}{\Lambda^2}\right), \eeq where
$\Lambda$ is a cut-off frequency. In the so-called Fokker-Planck
limit ($k_BT  \gg \hbar \Lambda$), we have \beq
\label{ohmiccase}
\nu(s)=\frac{2M_1k_BT\gamma}{\hbar}\delta(s),\,\,\, \eta(s)=M\gamma
\frac{d}{ds}\delta(s).\eeq Hence, $\delta\Omega^2=-2\gamma
\delta(0), \Gamma=\gamma, \Delta =0, \Sigma = 2M_1\gamma k_BT$. The
constant coefficients obtained for such a model give rise to a
Markovian master equation. The Wigner function for the center of
mass coordinates obeys the Fokker-Planck-Markov equation
\cite{footnote}:
\begin{eqnarray} \label{wignermarkov}
{\partial W_{\rm cm} \over\partial t}= &-&{P \over M_1}{\partial W_{\rm cm}\over
\partial X} -  M_1\Omega'^2 X^2 {\partial W_{\rm cm} \over \partial P}\nonumber \\
&+& 2\gamma \frac{\partial}{\partial P}(P W_{\rm cm}) \nonumber \\
&+&2M_1\gamma k_BT \frac{\partial^2}{\partial P^2} W_{\rm cm},
\end{eqnarray}
where $M_1=2M$ and $\Omega'^2=\Omega^2 +\delta\Omega^2$.

\section{The Influence Functional Method}
\label{propagatorsection}

In the last section we showed a simple derivation of the master
equation for the reduced density matrix and the Fokker-Planck
equation for the Wigner function. In general it is difficult to
get a general analytical solution of the master equation. It turns
out that  in some cases of interest, one can get analytic
solutions of the master equation through the influence functional
method \cite{AH}. Using this method, we can get the evolution
operator for the reduced density matrix or the evolution kernel
for the exact master equation which will be very useful for the
study of quantum decoherence and disentanglement problems.

Because of this, in this subsection, we will outline the key steps in
the derivation of the master equation (\ref{mastercoord}) via the
path integral method.

As before, the density matrix of the total system at any time $t$ can be written as
\begin{eqnarray}
 \rho(t) = e^{-i\frac{H_{\rm tot}t}{\hbar}} \rho(0) e^{i\frac{H_{\rm
 tot}t}{\hbar}}.
\end{eqnarray}
The reduced density matrix of the system is evolved by the
propagator $J_r$ from time $t=0$ to $t$ as
\begin{eqnarray}
\label{mainJ}
 \rho_r(x_1,x_2;y_1,y_2,t) &=& \int d q_n \langle x_1,x_2,q_n |
\rho(t) | y_1,y_2,q_n \rangle  \nonumber \\ &=& \int  dx_{0} dy_{0}
J_r(x_1,x_2,y_1,y_2,t;x_{10},
x_{20}, y_{10}, y_{20},0) \nonumber \\ && \times \rho_r(x_{10},
x_{20}; y_{10}, y_{20}; t=0),
\end{eqnarray}
where we have used the collective notation $dx_{0} dy_{0}= dx_{10}
dx_{20} dy_{10} dy_{20}$.

The evolution propagator $J_r$ can be written in a path-integral
representation as
\begin{eqnarray}\label{Joperator}
&&
J_r(x_1,x_2,y_1,y_2,t;x_1^{\prime},x_2^{\prime},y_1^{\prime},y_2^{\prime},0)
\nonumber \\ &=&\prod^2_{k=1} \int_{x_{ki}}^{x_{kf}} {\cal
 D}x_k\int_{y_{ki}}^{y_{kf}} {\cal
 D}y_k  \exp ( \frac{i}{\hbar} S_S[x_1,x_2]- \frac{i}{\hbar}
 S_S[y_1,y_2])
 \times  {\cal F}[x_1,x_2,y_1,y_2],
\end{eqnarray}
where $  {\cal F}[x_1,x_2,y_1,y_2] $ is the Feynman-Vernon influence
functional defined by
\begin{eqnarray}
{\cal F}[x_1,x_2,y_1,y_2] &=& \int
 dq_n^{\prime} d\tilde{q}_n^{\prime} d q_n   \rho_{bath}(q_n^{\prime} ,
\tilde{q}_n^{\prime}, 0)
 \int_{q_n^{\prime}}^{q_n} {\cal D}q_n  \int_{\tilde{q}_n^{\prime}}^{q_n} {\cal
 D}\tilde{q}_n
 \exp \{ \frac{i}{\hbar} (S_I[x_1,x_2,q_n]-\nonumber \\
&& S_I[y_1,y_2,\tilde{q}_n] + S_B[q_n] - S_B[\tilde{q}_n] ) \} \nonumber\\
 &=& \exp \{ \frac{i}{\hbar} (S_{IF} [x_1,x_2,y_1,y_2] ) \},
\end{eqnarray} where $S_{IF}$ is the influence action.
For the QBM model we are considering here, the influence action can
be written as:
\begin{eqnarray}
S_{IF}[x_1,x_2,y_1,y_2] &=&  - 2 \int_0^t d s_1 \int_0^{s_1} d
   s_2
   [\Delta_1(s_1) + \Delta_2(s_1)  ] \eta(s_1 - s_2)  [ \Sigma_1(s_2)  +
   \Sigma_2(s_2)]\nonumber \\
   &+  &  i \int_0^t d s_1  \int_0^{s_1} d
   s_2
   [\Delta_1(s_1) + \Delta_2(s_1)  ] \nu(s_1 - s_2)  [ \Delta_1(s_2)
   + \Delta_2(s_2)],
\end{eqnarray}
where
\begin{eqnarray}
\Sigma_1 = \frac{1}{2} (x_1 + y_1), \quad  \quad \Sigma_2
=\frac{1}{2} (x_2 + y_2), \quad  \quad \Delta_1 = x_1 -y_1, \quad
\quad \Delta_2 = x_2 -y_2.
\end{eqnarray}
Note that the integrand in Eq. (\ref{Joperator}) is Gaussian,
hence the integral can be computed exactly and the explicit form
of $J_r$ is,
\begin{eqnarray}
J_r &=& \tilde{N}\exp{(\frac{i}{2}S_I - S_R)},
\end{eqnarray}
where the expressions of $S_I$ and $S_R$ can be written in more
compact forms with the following notations: \beqa
x^+_k &=&x_{1k}+x_{2k}, \,\,\,\, y^+_k=y_{1k}+ y_{2k},\\
x^-_k&=&x_{1k}-x_{2k}, \,\,\,\, y^-_k=y_{1k}- y_{2k}, \eeqa whence
\begin{eqnarray}
S_I &=&  b_1 (x^+_t +y^+_t) (x^+_t -y^+_t) + b_2 (x^+_0 +y^+_0 )(x^+_t -y^+_t)
\nonumber \\
&-&  b_3 (x^+_t +y^+_t )(x^+_0 -y^+_0)-b_4 (x^+_0 +y^+_0 )(x^+_0 -y^+_0) \nonumber \\
&+&  b_5
(x^-_t +y^-_t )(x^-_t -y^-_t)+ b_6 (x^-_0+y^-_0 )(x^-_t -y^-_t)  \nonumber \\
&-&  b_7 (x^-_t +y^-_t )(x^-_0 -y^-_0)-b_8(x^-_0 +y^-_0 )(x^-_0 -y^-_0),
\end{eqnarray}
and
\begin{eqnarray}
S_R &=&  a_{11} (x^+_{t} - y^+_{t})^2 + a_{22}
(x^+_{0} - y^+_{0})^2 \nonumber \\
&&+
a_{12}(x^+_{0}-y^+_{0} )(x^+_{t} -
y^+_{t}).
\end{eqnarray}
The functions $b_i(t)$ and $a_{ij}(t)$ depend on the environment and
can be constructed from the solutions to the equations
\begin{eqnarray}
b_2(t) \equiv \frac{1}{2} \dot{u}_1(t), \quad b_1(t) \equiv
\frac{1}{2} \dot{u}_2(t), \quad b_6(t) \equiv \frac{1}{2}
\dot{w}_1(t), \quad b_5(t) \equiv \frac{1}{2} \dot{w}_2(t), \quad \nonumber \\
b_4(t) \equiv \frac{1}{2} \dot{u}_1(0), \quad b_3(t) \equiv
\frac{1}{2} \dot{u}_2(0), \quad b_8(t) \equiv \frac{1}{2}
\dot{w}_1(0), \quad b_7(t) \equiv \frac{1}{2} \dot{w}_2(0), \quad
\end{eqnarray}
where $w_i(t)$ are functions which satisfy the following equation
\begin{eqnarray}
\ddot{\bar{\Sigma}}(s) + \Omega^2 \bar{\Sigma}(s) =0,
\end{eqnarray}
with the boundary conditions:
\begin{eqnarray}
w_1(0)=1=w_2(t), \quad w_1(t) =0=w_2(0),
\end{eqnarray}
\begin{eqnarray}
a_{ij}(t)= \frac{1}{2} \int_0^t ds_1 \int_0^t ds_2 u_i(s_1)
\nu(s_1 -s_2 )u_j(s_2).
\end{eqnarray}

With the expression of $J_r$, we can derive the master equation for
the reduced density matrix (\ref{mastercoord}). This is shown in
Appendix A.


An exact form of the evolutionary operator for the reduced density
matrix is a priced object:  Not only can one  derive from it the
exact master equation for the reduced density matrix,  with this
explicit expression of the evolutionary operator, given any initial
reduced density matrix $\rho_r$ at time $t_0$ one can calculate
$\rho_r$ at any later time $t$ without having to solve the
complicated second order partial differential equation with
time-dependent coefficient functions.

For example, we will apply this evolutionary operator to the study of
the decoherence and disentanglement of two coupled harmonic
oscillators in a common heat bath. One can also use it to calculate
the higher moments of physical observables of interest such as the
position and the momentum operators which enter into the derivation
of a generalized uncertainty principle for composite objects at
finite temperature \cite{HZ2}. It can also be used to address the
issue of the influence of entanglement on the relation between the
statistical entropy of an open quantum system and the heat exchanged
with a low temperature environment such as studied in
\cite{HoeBue05}. Another interesting application would be the
entanglement between a qubit and an oscillator. Adopting a level
reduction scheme, Shiokawa and Hu \cite{ShiHu} used the evolutionary
operator of 1HO QBM to study the dynamics of the spin-boson model.
The explicit expression of the evolutionary operator for the 2HO QBM
may be used to construct effective 1HO-spin-boson models found in
many condensed matter quantum computer schemes for the analysis of
the interaction between a qubit and a harmonic oscillator and their
decoherence and disentanglement dynamics in the presence of a general
environment. See Section~\ref{summarysection} for a more detailed
exposition of further applications and extensions.

\section{Applications: Quantum Decoherence and Disentanglement, Uncertainty Relation for a Composite Object}
\label{application}
In this section we give three examples for the application of this
master equation: the decoherence and disentanglement of two coupled
harmonic oscillators in a common heat bath, and a derivation of the
uncertainty relation at finite temperature for a composite object
modeled by two harmonic oscillators in a general environment. For
some simplified cases we obtain analytic results which show
interesting features such as finite-time disentanglement
\cite{Yu-Eberly2004,Halliwell2004}.

\subsection{Dynamics of Quantum Coherence}
\label{subsec1}

We will assume that the system and the environment are initially
uncorrelated. The total density matrix at time $t=0$ then factorizes
into a product of density matrices for the system and the
environment. As usual, we further assume that the environment is initially in
thermal equilibrium at a given temperature $T$.

We assume initially the 2HO (labeled as 1 and 2) are separated with
distance $2L_0$ and the initial wave function of the 1-2 system  is
given by
\begin{eqnarray}
\Psi(x_1,x_2, t=0) &=&  s_1 \Psi_1(x_1) \Psi_1(x_2) + s_2 \Psi_1(x_1)\Psi_2(x_2)
\nonumber\\
&&+ s_3 \Psi_2(x_1) \Psi_1(x_2) + s_4 \Psi_2(x_1) \Psi_2(x_2),
\end{eqnarray}
where we have defined the displaced Gaussian states as
\begin{eqnarray}
\Psi_{1,2}(x) = N \exp{[-\frac{(x \mp L_0)^2}{2 \delta^2}]}
\exp{(\pm i P_0 x)},
\end{eqnarray}
and ${s_i}$ are any complex numbers subject to normalization
conditions. (We use 1,2 to label different initial positions of the
center of the Gaussian wave function of harmonic oscillators while
$x,y$ label  different time paths.)

With an initial reduced density matrix
\begin{eqnarray}
\rho_r(x_{10}, x_{20}; y_{10}, y_{20}; t=0) &=& \langle  x_{10},
x_{20} \mid \Psi(0) \rangle \langle \Psi(0) \mid y_{10}, y_{20}
\rangle \nonumber \\ &\equiv& \sum_{i,j} s_i s_j^* \rho_{ij}(x_{10},
x_{20}; y_{10}, y_{20}; t=0),
\end{eqnarray}
the reduced density matrix at $t$ is given by
\begin{eqnarray}
\label{influencefunctional} \rho_r(x_{1}, x_{2}; y_{1}, y_{2}; t)
&=& \int dx_{0} dy_{0} J_r(x_1,x_2,y_1,y_2,t;x_{10}, x_{20}, y_{10},
y_{20},0) \nonumber \\ && \times \rho_r(x_{10}, x_{20}; y_{10},
y_{20}; t=0).
\end{eqnarray}
Because the QBM model is linear and the initial state is Gaussian,
we can solve the master equation exactly for the dynamics of the
2HO system interacting with an environment with a general spectral
density at any temperature. Therefore, we can obtain the total
density matrix if the explicit solutions for each component are
known,
\begin{eqnarray} \rho_{ij}(x_{1}, x_{2}; y_{1}, y_{2}; t) &=& \int
dx_{0} dy_{0}
J_r(x_1,x_2,y_1,y_2,t;x_{10}, x_{20}, y_{10}, y_{20},0) \nonumber \\
&& \times \rho_{ij}(x_{10}, x_{20}; y_{10}, y_{20}; t=0).
\end{eqnarray}

Note that since $J_r $ and $\rho_{ij}$ are in the form of an
exponential with an exponent which is a quadratic function in
$(x_{10}, x_{20}; y_{10}, y_{20})$, we can use a standard trick for
the evaluation of the integral,
\begin{eqnarray}
\label{decoherence1}
\rho_{ij}( t) &=& \int  dx_{0} dy_{0} J_t \times \rho_{ij}( t=0) \nonumber \\
&=& \int   dx_{0} dy_{0} \exp{[- \vec{x}^T \cdot G_{ij} \cdot
\vec{x} + \frac{1}{2} \vec{F}_{ij}^T \cdot \vec{x} + \frac{1}{2}
\vec{x}^T \cdot \vec{F}_{ij} + c_{ij} ]} \nonumber \\ &=&
\frac{(\sqrt{\pi})^4}{\sqrt{\det{G_{ij}}}}\exp{(c_{ij} +
\frac{1}{4}\vec{F}^T_{ij} \cdot G_{ij}^{-1} \cdot \vec{F}_{ij})},
\end{eqnarray}
where $\vec{x}^T = (x_{10}, x_{20}, y_{10}, y_{20})$.

Once we have $\rho_{ij}(x_{1}, x_{2}; y_{1}, y_{2}; t) $ we can
perform the following substitution $x_1 \mapsto X_1 -\frac{z_1}{2} ;
x_2 \mapsto X_2 -\frac{z_2}{2} ; y_1 \mapsto X_1 +\frac{z_1}{2} ; y_2
\mapsto X_2 +\frac{z_2}{2}  $ and then do the Fourier transform to
get the Wigner function at a later time $t$:
\begin{eqnarray}
\label{decoherence2}
W_{ij}(X_1,X_2,P_1,P_2,t) &=& \int \int \frac{dz_1 dz_2}{(2 \pi
\hbar)^2} \exp{(i P_1 z_1 + i P_2 z_2)}\nonumber \\
&&\times \rho_{ij}(X_1 -\frac{z_1}{2},X_2 -\frac{z_2}{2} ; X_1
+\frac{z_1}{2}, X_2 +\frac{z_2}{2}; t).
\end{eqnarray}
Since after the substitution the exponent of $\rho_{ij}$ is
quadratic in $z_1, z_2$, the above integration can be evaluated
explicitly. These solutions (\ref{decoherence1}) and
(\ref{decoherence2})  will be useful in  decoherence and
disentanglement analysis below. The detailed results and the
explicit expressions  of $\rho_{ij}$ can be found in Appendix B.

When viewed from the center of mass coordinate the physics of
decoherence for a 2HO system is essentially similar to that described
in \cite{HPZ,PHZ} using the Hu-Paz-Zhang master equation for 1HO
because the environment couples to the system only through the center
of mass coordinate $X$ and is independent of the relative coordinate
$x$. The evolution of the relative coordinate part in the reduced
density matrix is unitary and hence will not affect the decoherence
processes. One can easily recognize these features from
(\ref{masterosci}) and (\ref{relation}). The effects of
environment-induced decoherence are encoded in the coefficient
functions $a(t), b(t), c(t), d(t)$ of (\ref{masterosci}). As one can
see from this example  four of the matrix elements $\rho_{11},
\rho_{14},\rho_{41}, \rho_{44}$ are similar to those in the example
considered in  \cite{PHZ} sans the relative coordinates.

However, the issue of disentanglement is quite different because
usually the entanglement measure is related to the global property of
the whole reduced density matrix.  In general, entanglement involves
both the center of mass and the relative coordinate dynamics. It is
difficult to make any prediction on how disentanglement evolves from
the information of only the 1HO system. For instance, while the cm
coherence always disappear asymptotically, in contrast, entanglement
of the two particles may terminate in a finite time. In the third
subsection, we will address this issue with a simple illustrative
example.

\subsection{Uncertainty Principle for Composite Objects}

In this subsection, the generalized uncertainty relation for a
composite object is investigated from the viewpoint of quantum open
systems. Here the system is modeled by two harmonic oscillators and
the environment by a heat bath at temperature $T$. As such, both
thermal fluctuation and quantum noise come to play when the
uncertainty relation between position and momentum is considered
\cite{HPZ,HZ2}.

The exact solution for the two harmonic oscillators coupled to a
common heat bath can be found by decomposing the total system into
two fictitious surrogate subsystems, namely, the subsystems described
by the center of mass and the relative coordinates, respectively.
Such a decomposition guarantees that the two subsystems are
decoupled, and as such, the solution of the total system is a tensor
product of the two subsystems: \beq \rho_r=\rho_{\rm cm}\otimes
\rho_{\rm rel}. \eeq Using the center of mass coordinate as described
by the Hamiltonian (\ref{tot}), the complete information about the
state of the open system is contained in the reduced density operator
$\rho_r(t)$.

For a class of initial Gaussian states given by \beq \psi(x,0)= {N_0}
\exp\left[ -\frac{(x-x_0)^2}{4\sigma^2} +\frac{i}{\hbar}p_0 x\right]
\eeq where, ${N_0}=1/({2\pi \sigma^2})^{\frac{1}{4}}$,  the initial
density operator for each fictitious harmonic oscillator in the
coordinate representation can be written as: \beq \rho(x,x', 0)=
\psi^*(x,0)\psi(x',0)=N^2_0 \exp\left[ -\frac{(x-x_0)^2}{4\sigma^2}
-\frac{(x'-x_0)^2}{4\sigma^2} -\frac{i}{\hbar}p_0 x
+\frac{i}{\hbar}p_0 x'\right]. \eeq In order to compute the variance
of position and momentum operators, it is more convenient to use the
Wigner function which can be obtained from the Fourier transform of
the density operators (\ref{wignerdensity}).  To be more specific,
for the harmonic oscillator representing the center of mass degree of
freedom, the corresponding Wigner function is simply given by: \beq
W_{\rm cm}(X, P)=N^2_0 \exp\left[ -\frac{(X-x_0)^2}{2\sigma^2}
-\frac{2\sigma^2}{\hbar^2} (P-p_0)^2 \right] \eeq The
variance of the operator $X, P$ denoted  by $(\Delta X)^2=\langle
X^2\rangle - \langle X \rangle^2 $ and $(\Delta P)^2=\langle
P^2\rangle - \langle P \rangle^2$ can be computed easily \beqa
\langle X^2\rangle &=& \frac{1}{2\pi \hbar} \int dXdP  X^2 W_{\rm cm}(X,P,t)\\
 \langle P^2\rangle &=&  \frac{1}{2\pi \hbar}  \int dXdP  P^2 W_{\rm cm}(X,P,t)
\eeqa
 where $W_{\rm cm}(X,P,t)$ is the solution of the Fokker-Planck equation for
a single harmonic oscillator (see Appendix B or \cite{HZ2}). In
particular, for an ohmic environment (\ref{ohmiccase}), the
uncertainly relation in the weak  damping limit ($\gamma  <<
2\Omega$) is given by \beq \label{unrelation} U(t)=(\Delta
X)^2(\Delta P)^2(\Delta x)^2(\Delta p)^2=f_{\rm cm}(t)f_{\rm rel}(t)
\eeq with
 \beqa
f_{\rm cm}(t) &=&\frac{\hbar^2}{4}\left[ e^{-\gamma t} + \coth  \frac{\hbar \Omega'}{2k_B T} (1- e^{-\gamma t})\right]^2\nonumber\\
&+& \hbar^2 \coth  \frac{\hbar \Omega'}{2k_B T} \left[ \frac{(1-\delta)^2}{4\delta}\left(1-e^{-\gamma t}\right) -\frac{(1-\delta^2)\gamma}{8 \Omega' \delta } \sin  2\Omega' t \right]e^{-\gamma t}\nonumber\\
&+&\hbar^2 \left[ \frac{1-\delta^2}{4\delta} \sin  2\Omega' t  +\frac{\gamma}{2\Omega'} \left( \coth  \frac{\hbar \Omega'}{2k_B T} - \frac{1+\delta^2}{2\delta}  \right)  \sin^2  \Omega' t  \right]^2 e^{-2\gamma t},\\
f_{\rm rel}(t)  &=&  \frac{\hbar^2}{4}\left[
1+\frac{1}{4\delta^2}(1-\delta^2)^2\sin^2 2\Omega t \right]. \eeqa
where $\Omega'=\sqrt{\Omega^2-\gamma^2/4}$ and
$\delta=2\Omega\sigma^2/\hbar$.
At short times ($ t << 1/\gamma, 1/\Omega$), \beqa
f_{\rm cm}(t) &=& \frac{\hbar^2}{4}\left[ 1+ 2(\delta \coth \frac{\hbar \Omega'}{2k_B T} -1)\gamma t  \right] \\
f_{\rm rel}(t)  &=&  \frac{\hbar^2}{4} .\eeqa In this short time span,
the time-dependent quantum dispersion of the wave packet constructed
in the relative coordinate may be ignored. It is interesting to
compare the uncertainty relation (\ref{unrelation}) with that between
the $x_1,x_2$ and $p_1,p_2$, denoted by $U_{\rm x_ip_i}$: \beq
\label{unrelation1} U_{\rm x_ip_i}=(\Delta x_1)^2(\Delta
p_1)^2(\Delta x_2)^2(\Delta p_2)^2 \geq \frac{1}{8} U(t). \eeq As will
be shown in the next subsection, the variance of the operators $x$
and $X$ etc can indeed provide some useful information about the
evolution of quantum entanglement of the Gaussian states.

\subsection{Dynamics of Entanglement: An Example}

As shown in Subsection \ref{subsec1}, the decoherent effects of a
thermal heat bath is captured by the influential functional appearing
in (\ref{influencefunctional}).  An environment that destroys quantum
coherence can also disentangle two quantum Brownian particles. The
dynamics of decoherence and entanglement of two harmonic oscillators
interacting with a common environment is useful for understanding
some basic issues in macroscopic quantum phenomena. We will present a
more detailed study of this issue in a later paper. Here we show a
simple example which has analytic solutions. Take as initial state
the Wigner function: \beq \label{wignerexample}
W(x_1,x_2,P_1,P_2)=W_{\rm cm}(X,P)W_{\rm
rel}(x,p)=e^{-\frac{X^2}{2a^2}-\frac{P^2}{2b^2}
}e^{-\frac{x^2}{2c^2}-\frac{p^2}{2d^2}}. \eeq where $P, X, x$ and $p$
are canonical variables defined in (\ref{trans1}) and (\ref{trans2}).
We have omitted an irrelevant normalization factor. Note that the
widths $a^2, b^2, c^2$ and $d^2$ cannot be chosen arbitrarily since
they have to satisfy the uncertainty relations: \beq a^2b^2\geq
\frac{\hbar^2}{4}, \,\,\, c^2d^2\geq \frac{\hbar^2}{4}. \eeq For a
wide range of parameters $a,b,c$ and $d$, the Wigner function
$W(X,P,x,p)$ is entangled, since generally it cannot be written as a
product of $W_1(x_1,P_1)$ and $W_2(x_2,P_2)$. At any time $t$, it is
known that the separability of the state (\ref{wignerexample}) can be
easily detected \cite{duan,simon}.

Now we consider the dynamics of this state under the influence of a
common environment. For greatest simplicity, we assume two free
particles coupled to a Markovian thermal bath (Setting $\Omega=0$
and $\kappa =0$) and assume the dissipation in cm coordinates is
negligible. Under these conditions, the Wigner equation $W_{\rm
cm}(X,P)$ for cm coordinates (\ref{wignermarkov}) takes on a simple
form: \beq \frac{\partial W_{\rm cm}}{\partial
t}=-\frac{P}{M_1}\frac{\partial W_{\rm cm}}{\partial X} +
 D\frac{\partial^2 W_{\rm cm}}{\partial P^2},
\eeq where $D=2M_1 \gamma k_BT$. The solution for the dissipative
evolution of the center of mass can be easily obtained, and from it,
we can compute the variances  of $X$ and $P$ at time $t$ to be: \beqa
(\Delta X^2)(t)&=&\frac{2Dt^3}{3M^2}  +\frac{b^2t^2}{4M^2}  +a^2,\\
(\Delta P^2)(t)&=& 2Dt+b^2. \label{variance1} \eeqa  Since the
evolution of the Wigner function $W_{\rm rel}(x,p)$ for the relative
coordinates $x, p$ is unitary, \beq \frac{\partial W_{\rm
rel}}{\partial t}=-\frac{p}{M_2}\frac{\partial W_{\rm rel}}{\partial
x}. \eeq the variances at $t$ are simply given by \beqa
\label{variance2}
(\Delta x^2)(t) &=&  \frac{4 d^2}{M^2}t^2 +c^2,\\
(\Delta p^2)(t) &=& d^2. \eeqa According to \cite{duan},  we may
choose the EPR-like operators as : \beq u={\tilde x}_1 -{\tilde
x}_2, v={\tilde P}_1 + {\tilde P}_2,\eeq where $ {\tilde x}_i,
{\tilde P}_i\,\,\,(i=1,2)$ are the dimensionless variables
satisfying $[{\tilde x}_i, \,{\tilde P}_j]=i \delta_{ij}$, \beq
{\tilde x}_i=\left(\frac{M D}{\hbar^3}\right)^{\frac{1}{4}}x_i,
\,\,\, {\tilde P}_i=\left(\frac{1}{\hbar M
D}\right)^{\frac{1}{4}}P_i, \,\,\ (i=1,2). \eeq Then the Gaussian
state (\ref{wignerexample}) at $t$ is disentangled if
 and only if the following inequality is satisfied
\beq \label{sufficient} (\Delta u^2)(t)+ (\Delta v^2)(t)  \geq 2.
\eeq Inserting (\ref{variance1}) and (\ref{variance2}) into the
above inequality, one gets, \beq \label{inequality} At^2+Bt +C\geq
2, \eeq where \beqa
A &=& \frac{4d^2}{M^2}\sqrt{\frac{MD}{{\hbar^3}}},\\
B &=& 2\sqrt{\frac{D}{\hbar M}},\\
C &=& \frac{b^2}{\sqrt{\hbar M D}}+c^2\sqrt{\frac{MD}{{\hbar^3}}}.
\eeqa From (\ref{inequality}), the disentanglement time $t_{\rm
dent}$ can be determined to be \beq \label{dtime} t_{\rm
dent}=\frac{-B+\sqrt{B^2-4AC+8A}}{2A}. \eeq Thus after $t \geq
t_{\rm dent}$  the state (\ref{wignerexample}) becomes completely
separable.

In situations when the 2HO are coupled or share the same environment,
it is expected that for some initial states entanglement will persist
longer than the case when there is no direct coupling between the two
oscillators and each of them is coupled to a separate environment
(See, e.g. \cite{ASH} for two qubits in a common electromagnetic
field). This is what one might anticipate would happen for our model
in the more general cases. On the other hand, as shown in this
simplified example, finite-time disentanglement may yet occur for
some initial states when there is no direct coupling between the two
oscillators.

Such finite-time decay behavior has been noted before in several
cases where two qubits \cite{Yu-Eberly2004} or two harmonic
oscillators \cite{Halliwell2004,LD} are individually coupled to
their own heat baths. We show here the onset of  the finite-time
decay for the case of a common heat bath.  However, it should be
emphasized again that the finite-time disentanglement process
found here depends crucially on the choice of initial states
because for some initial states the mutual actions between the two
harmonic oscillators may lead to entanglement generation.  As
shown in the case of two-qubits under phase noises, when the
initial states are protected by a decoherence-free subspace
quantum entanglement is shown to be robust against the thermal
noise \cite{Yu-Eberly2002}. The 2HO model considered here will
exhibit similar features, but further details  will go beyond the
scope of this paper.

\section{Further Applications and Developments}
\label{summarysection}

\paragraph{Summary} In this work we derive an exact master equation
for two coupled quantum harmonic oscillators interacting via bilinear
coupling with a common environment at arbitrary temperature made up
of many harmonic oscillators with a general spectral density
function. We first show a simple derivation based on the observation
that the two harmonic oscillator model can be effectively mapped into
that of a single harmonic oscillator in a general environment plus a
free harmonic oscillator. Since the exact one harmonic oscillator
master equation is available \cite{HPZ} the exact master equation
with all its coefficients for this two harmonic oscillator model can
be easily deduced from the known results of the single harmonic
oscillator case.  In the second part we give an influence functional
treatment of this model and provide explicit expressions for the
evolutionary operator of the reduced density matrix which are useful
for the study of decoherence and disentanglement issues. We show
three applications of this master equation: on the decoherence and
disentanglement of two harmonic oscillators due to their interaction
with a common environment  and a derivation of the uncertainty
principle at finite temperature for a composite object, modeled by
two interacting harmonic oscillators. For the example of entanglement
dynamics under Markovian approximation we find finite-time
disentanglement taking place for a Gaussian state.

\paragraph{Decoherence and Disentanglement}
We mention some further developments and applications where our
analysis of the 2HO QBM model can be usefully extended to or compared
with. First, for the study of decoherence and disentanglement between
two observers,  a direct comparison can be carried out with some
recent findings in \cite{LCH} where the model of two harmonic
oscillators in relativistic motion (one could be in uniform
acceleration) in a common field in Minkowsky or a black hole
spacetime. In the latter situation it is of interest to see how
entanglement and teleportation will be affected by its unusual causal
properties. The case of two oscillators in inertial motion in
ordinary Minkowsky spacetime would correspond to our problem here
after invoking Lorentz invariance. Second, pursuant to our analysis
of the uncertainty principle for composite objects, the substance of
our calculations there could be applied to another interesting
physical issue pertaining to the Landauer principle \cite{Landauer}
and the Clausius inequality. Landauer principle which rests at the
foundation of the thermodynamics of information processing, states
that (paraphrased in the words of Bennett \cite{Bennett}) ``any
logically irreversible manipulation of information, such as the
erasure of a bit or the merging of two computation paths, must be
accompanied by a corresponding entropy increase in non-information
bearing degrees of freedom of the information processing apparatus or
its environment. Conversely, it is generally accepted that any
logically reversible transformation of information can in principle
be accomplished by an appropriate physical mechanism operating in a
thermodynamically reversible fashion". (See also
\cite{Cav93,Bub,Maroney}, the last contains a proposal for a
generalized Landauer's principle.) It is well known that the root of
this relation is the second law of thermodynamics, but how to measure
a logical operation in physical terms or to associate a logical state
or its transformation with an energy cost and an entropy increase is
a new challenge.

\paragraph{Quantum Information and Thermodynamics}
There are many angles to see how Landauer's bound in quantum
information theory is related to Clausius' inequality in classical
thermodynamics. One such approach is by way of quantum open systems
which can treat the dynamics of the system and its quantum
information content in fully nonequilibrium settings. This is the
conceptual framework and technical systematics we have adopted. Here,
dissipation and decoherence in the system and disentanglement between
the system and its environment may be followed closely by the
evolution of the reduced density matrix (RDM), and the entropy change
of the system in the thermodynamic limit may be calculated, with
little difficulty. In this vein, using the quantum Brownian model
(QBM) of the Caldeira-Leggett (CL)  type Hoerhammer and Buettner
\cite{HoeBue05} investigated the influence of entanglement on the
relation between the statistical entropy of an open quantum system
and the heat exchanged with a low temperature environment. (See also
\cite{HoeBue07}). Their two Brownian oscillator model is of
particular relevance to our work here. Compared to the case of a
single Brownian particle, two coupled harmonic oscillators can
account for how the internal degrees of freedom of the system would
affect the heat and entropy changes. Because they adopted the CL
treatment their results are subcases of ours here  (in the same way
that the CL treatment \cite{CalLeg} of QBM is related to the HPZ
treatment \cite{HPZ}, viz, the latter preserves the positive
definiteness of RDM in its entire evolution and the HPZ master
equation extends the range of validity to non-Markovian regimes.) The
CL results are valid only for ohmic baths at high temperatures
pertaining to the Markovian regime. For low temperatures and nonOhmic
baths pertaining to the nonMarkovian regimes the HPZ treatment is
expected to yield more accurate results. Thus using the master
equation presented here for the 2HO QBM model following HPZ treatment
and the analytical solutions found recently \cite{FleRouHu}  for
various parameter ranges one could obtain an improved Landauer bound
for quantum information processing in the nonMarkovian regimes. On
the other side of the balance, the Clausius inequality, operative
only in the thermodynamic limit, would be too coarse a measure for
the energy cost and entropy change of quantum information processing
anyway. With the master equations derived here there is much room for
tightening the Landauer bound.

\paragraph{Qubit - Oscillator Entanglement} As subcases of the present
model one can investigate the interaction between a two-level system
with a harmonic oscillator in a general environment which is of
general interest for quantum computer design purposes. One could
apply a level reduction scheme such as that used in \cite{ShiHu} to
one of the two harmonic oscillators, turning the 2HO-bath model into
an effective 1HO-spin-boson model where the bimodal oscillator mocks
up a qubit. The simpler case without an environment would correspond
to a two level atom in a multi-mode cavity, such as studied in
\cite{CumHu}. Doing a level reduction scheme for both oscillators and
viewing the harmonic oscillator bath as a field would reduce our 2HO
QBM model to that of two qubits interacting either directly or
indirectly through a common field. An example of the latter situation
is studied in \cite{ASH}. One can use the exact master equations here
under appropriate simplifications to describe the nonMarkovian
dynamics of such systems.

\paragraph{Quantum Superposition of Two Mirrors}
As mentioned in the Introduction, a new category of problems which
has received much attention lately is represented by the quantum
superposition of two mirrors \cite{Marshall}. The two mirrors can be
modeled by two quantum harmonic oscillators, but in most models for
proposed experimental designs, the mirrors are coupled by radiation
pressure.  This class of models with photon number - mirror
displacement (Nx) type of coupling used for mirror-photon
entanglement \cite{Vitali1}, entanglement cooling of a mirror
\cite{Vitali2} and entanglement of test masses and standard quantum
limit \cite{EntSQL} is very different from the class with bilinear
coupling in QBM studies (Beware of inconsistencies in the usual
master equations for this problem, see \cite{YuFleHu}). On the
surface the convenience of the 2HO model which possesses many useful
solutions would not be readily available, but a recent observation by
Galley could provide a bridge to these two common classes of models
and unleash the resources gathered from the 2HO QBM problem for the
solution of this type of quantum optics problems. (See
\cite{GalChoHu}.).

\paragraph{Macroscopic Quantum Phenomena}
Finally,  a whole range of issues in macroscopic quantum phenomena
can be addressed with the master equation (or the associated Langevin
or Fokker-Planck equations) derived here. In particular, decoherence
and disentanglement in 2HO system under more general conditions and
$N$-harmonic oscillators systems \cite{CHY07} are currently under
study. It can also be applied to the analysis of quantum decoherence,
entanglement, fluctuations, dissipation and teleportation of electro-
opto-mechanical systems and superposition of moving mirrors due to
quantum and radiative effects.

\section*{Acknowledgements}
C.-H. C. would like to thank Dr. Kazutomu Shiokawa for discussions
on using effective spin-boson model in the treatment of quantum
entanglement. T. Y. would like to thank Prof. J. H. Eberly for
many useful conversations and acknowledges support  from ARO Grant
W911NF-05-1-0543 to the University of Rochester. BLH is partially
supported by the NSF (PHY-0426696) under the ITR program and by
NSA-LPS to the University of Maryland. Part of this work was done
while we enjoyed the hospitality of the Institute of Physics of
the Academia Sinica, Taipei, the National Center for Theoretical
Sciences and the Center for Quantum Information Sciences at the
National Cheng Kung University, Tainan, Taiwan.

\begin{appendix}
\label{final}

\section{Derivation of Exact  Master Equation From Path Integral}
Deriving the master equation from the path integral is lengthy,
but one of the advantages of this derivation is that the explicit
form of the propagator can be used to find an explicit solution of
the equation in many interesting cases.  We will mainly  follow
the steps in \cite{HPZ} and outline the key steps in deriving the
master equation from the path integral method.

From (\ref{mainJ}), it is easy to see that, to get the master
equation, one first needs to calculate $J_r(t+dt,0) - J_r(t,0)$.
The complete derivation can be decomposed into the following four
steps.

\subsection{Step one}
Our first task is to take the functional representation of
$J_r(t+dt,0)$ and divide each of the path integrals into two parts.
We introduce four intermediate points $x_{1m}, x_{2m}, y_{1m},
y_{2m}$ at time $t$ and integrate over them, thus symbolically, we
write
\begin{eqnarray}
\int_{0;x_{10}}^{t+dt;x_{1f}} {\cal D}x_{1} \int_{-\infty}^{\infty}
dx_{1m} \int_{0;x_{10}}^{t;x_{1m}} {\cal D}\bar{x}_{1}
\int_{t;x_{1m}}^{t+dt;x_{1f}} {\cal D}\tilde{x}_{1}.
\end{eqnarray}
There are three similar expressions for the sum over $x_{2}, y_{1},
y_{2}$ histories.

The original histories $x_1(\tau)$ are functions defined on
$(0,t+dt)$ time interval with $x_1(0) = x_{10}, x_1(t+dt)=
x_{1f}$. The new set of histories $\bar{x}_1(\tau),
\tilde{x}_1(\tau)$ are functions defined on $(0,t), (t, t+dt)$
intervals with $\bar{x}_1(0)=x_{10}, \bar{x}_1(t)= x_{1m},
\tilde{x}_1(t)= x_{1m}, \tilde{x}_1(t+dt)= x_{1f}.$

So we can write
\begin{eqnarray}
A[x_1,x_2,y_1,y_2] &=&  S_S[x_1,x_2] - S_S[y_1,y_2]
 + \delta A[x_1,x_2,y_1,y_2] \nonumber
 \\ &=& A[\bar{x}_1, \bar{x}_2, \bar{y}_1,
\bar{y}_2] + A[\tilde{x}_1, \tilde{x}_2, \tilde{y}_1, \tilde{y}_2] +
A_i[\bar{x}_1, \bar{x}_2, \bar{y}_1, \bar{y}_2, \tilde{x}_1,
\tilde{x}_2, \tilde{y}_1, \tilde{y}_2],
\end{eqnarray}
where $A_i$ term mixes the $\tilde{x}$ histories with $\bar{x}$
ones. The appearance of the $A_i$ term is due to the non-locality of
the influence functional.

\subsection{Step two}
Next, we will use straight line histories approximation of
$(\tilde{x}_1, \tilde{x}_2, \tilde{y}_1, \tilde{y}_2).$ First, note
that
\begin{eqnarray}
\tilde{x}_1(s) = x_{1m} + (x_{1f} - x_{1m}) \frac{s-t}{dt} \equiv
x_{1m} + \beta_{1x} \frac{s-t}{dt} ,
\end{eqnarray}
and similarly,
\begin{eqnarray}
\tilde{x}_2(s) = x_{2m} + (x_{2f} - x_{2m}) \frac{s-t}{dt} \equiv
x_{2m} + \beta_{2x} \frac{s-t}{dt},
\end{eqnarray}
\begin{eqnarray}
\tilde{y}_1(s) = y_{1m} + \beta_{1y} \frac{s-t}{dt} , \quad
\tilde{y}_2(s) = y_{2m} + \beta_{2y} \frac{s-t}{dt}.
\end{eqnarray}
To compute the time derivative of $J_r$,  take the limit $dt
\rightarrow 0$. Thus we can write
\begin{eqnarray}
 & & \prod^2_{k=1}\int_{0;x_{k0}}^{t+dt;x_{kf}} {\cal D}x_{k}
\int_{0;y_{k0}}^{t+dt;y_{kf}} {\cal D}y_{k} \exp{(\frac{i}{\hbar}
A_[x_1,x_2,y_1,y_2 ])}\nonumber \\ &=& N(t)   \prod^2_{k=1} \int_{-\infty}^{\infty}
dx_{km}
dy_{km} \exp{(\frac{i}{\hbar}
A[\tilde{x}_1, \tilde{x}_2, \tilde{y}_1, \tilde{y}_2] )} \nonumber\\
& \times &
\prod^2_{k=1} \int_{0;x_{k0}}^{t;x_{km}} {\cal D}\bar{x}_{k}
\int_{0;y_{k0}}^{t;y_{km}} {\cal D}\bar{y}_{k}
\exp{(\frac{i}{\hbar} A[\bar{x}_1, \bar{x}_2, \bar{y}_1,
\bar{y}_2] )} \exp{(\frac{i}{\hbar} A_i[\bar{x}_1, \bar{x}_2,
\bar{y}_1, \bar{y}_2, \tilde{x}_1, \tilde{x}_2, \tilde{y}_1,
\tilde{y}_2] )}.
\end{eqnarray}

Expanding $A$  in $dt$ and keeping the contributions of the first
order terms, we get,
\begin{eqnarray}
A[\tilde{x}_1, \tilde{x}_2, \tilde{y}_1, \tilde{y}_2]  \approx
\frac{m}{2dt} (\beta_{1x}^2 + \beta_{2x}^2 -\beta_{1y}^2
-\beta_{2y}^2 ) -\frac{1}{2} m \Omega^2 dt (x_{1f}^2 + x_{2f}^2
-y_{1f}^2 -y_{1f}^2 ) + \cdot \cdot \cdot,
\end{eqnarray}
and
\begin{eqnarray}
 A_i[\bar{x}_1, \bar{x}_2, \bar{y}_1,
\bar{y}_2, \tilde{x}_1, \tilde{x}_2, \tilde{y}_1, \tilde{y}_2]
&\approx&  - dt \int_{0}^{t} d s J_{\vec{\Sigma}}(s) (
\bar{\Sigma}_1(s) + \bar{\Sigma}_2(s)  ) \nonumber \\
&&+ i dt \int_{0}^{t} d s J_{\vec{\Delta}}(s) ( \bar{\Delta}_1(s) +
\bar{\Delta}_2(s)),
\end{eqnarray}
where
\begin{eqnarray}
J_{\Sigma_1} + J_{\Sigma_2} &\equiv&  J_{\vec{\Sigma}}(s)  \frac{2}{dt}
\int_{t}^{t+dt} d s^{\prime} (
\tilde{\Delta}_1(s^{\prime}) +  \tilde{\Delta}_2(s^{\prime})  )
\eta (s^{\prime} - s) \nonumber \\
&\approx&  2( x_{1f} - y_{1f} +  x_{2f} - y_{2f} )\eta (t - s) +
\cdot \cdot \cdot,
\end{eqnarray}
and
\begin{eqnarray}
J_{\Delta_1} + J_{\Delta_2}  &\equiv&  J_{\vec{\Delta}}(s)  \frac{1}{dt}
\int_{t}^{t+dt} d s^{\prime} (
\tilde{\Delta}_1(s^{\prime}) + \tilde{\Delta}_2(s^{\prime})  ) \nu
(s^{\prime} - s) \nonumber \\
&\approx&  ( x_{1f} - y_{1f} +  x_{2f} - y_{2f} )\nu (t - s)  +
\cdot \cdot \cdot.
\end{eqnarray}
Here we can  keep  only terms up to the first order in
$\beta_i^2$.

In summary, the propagator $J_r$ can be formally written as
\begin{eqnarray}
& & J_r(x_{1f}, x_{2f}, y_{1f}, y_{2f}, t+dt | x_{10}, x_{20},
y_{10}, y_{20}, 0 ) \\ & \approx &  N(t)  \int_{-\infty}^{\infty}
d \beta_{1x}
 \int_{-\infty}^{\infty} d \beta_{2x}
  \int_{-\infty}^{\infty} d \beta_{1y}
   \int_{-\infty}^{\infty} d \beta_{2y} \exp{ ( \frac{i m}{2 \hbar dt} (\beta_{1x}^2
+ \beta_{2x}^2 -\beta_{1y}^2
-\beta_{2y}^2 ) )  } \\
& \times & \{ 1 - \frac{i}{\hbar} dt [V(x_{1f}, x_{2f}) -
V(y_{1f}, y_{2f})  ] \} \nonumber \\
&&\times  \tilde{J}_r(x_{1m}, x_{2m}, y_{1m}, y_{2m}, t+dt | x_{10},
x_{20}, y_{10}, y_{20}, 0 ; [\vec{b}] ),
\end{eqnarray}
where
\begin{eqnarray}
& &  \tilde{J}_r(x_{1m}, x_{2m}, y_{1m}, y_{2m}, t+dt | x_{10},
x_{20}, y_{10}, y_{20}, 0 ; [\vec{b}] ) = \\
& &  \int_{0;x_{10}}^{t;x_{1m}} {\cal D}\bar{x}_{1}
\int_{0;x_{20}}^{t;x_{2m}} {\cal D}\bar{x}_{2}
\int_{0;y_{10}}^{t;y_{1m}} {\cal D}\bar{y}_{1}
\int_{0;y_{20}}^{t;y_{2m}} {\cal D}\bar{y}_{2} \exp{(\frac{i}{\hbar}
A[\bar{x}_1, \bar{x}_2, \bar{y}_1, \bar{y}_2]   )} \\ &&
\exp{[\frac{i}{\hbar}(  - dt \int_{0}^{t} d s J_{\vec{\Sigma}}(s) (
\bar{\Sigma}_1(s) + \bar{\Sigma}_2(s)  ) + i dt \int_{0}^{t} d s
J_{\vec{\Delta}}(s) ( \bar{\Delta}_1(s) + \bar{\Delta}_2(s)  ) )]},
\end{eqnarray}
and
\begin{eqnarray}
\vec{b} = \left(%
\begin{array}{c}
  J_{\vec{\Sigma}} \\
  J_{\vec{\Delta}} \\
\end{array}%
\right),
\end{eqnarray}
where the sources $ \vec{b} $ are functions of the end points.
Note that $\tilde{J}_r(\vec{b})$ can be interpreted as the
evolution operator under the action of two external sources.

\subsection{Step three} Computation of the path integral
$\tilde{J}_r(\vec{b})$ can be done as follows. First, one can show
that
\begin{eqnarray}
& &  \tilde{J}_r(x_{1m}, x_{2m}, y_{1m}, y_{2m}, t | x_{10},
x_{20}, y_{10}, y_{20}, 0 ; [\vec{b}] ) = \\
& &  J_r(x_{1m}, x_{2m}, y_{1m}, y_{2m}, t | x_{10}, x_{20}, y_{10},
y_{20}, 0 )W(x_{1m}, x_{2m}, y_{1m}, y_{2m}, x_{10}, x_{20}, y_{10},
y_{20}, dt ).
\end{eqnarray}
(Note that the function $J_r$ is the evolution operator without
source while the function $W$ is a function of the end points. )

Then  we may parametrize the paths, and write
\begin{eqnarray}
& & \Sigma_1(s) = \varphi_1(s) + \Sigma_{cl,1}(s), \quad
\Sigma_2(s) = \varphi_2(s) + \Sigma_{cl,2}(s) \\ & & \Delta_1(s) \psi_1(s) +
\Delta_{cl,1}(s) , \quad \Delta_2(s) = \psi_2(s) +
\Delta_{cl,2}(s)
\end{eqnarray}
where the "classical paths" $\left(%
\begin{array}{c}
  \Sigma \\
  \Delta \\
\end{array}%
\right)_{cl}$ are the solutions to the equation of motion derived
from the real part of $A[\Sigma_1, \Sigma_2, \Delta_1, \Delta_2]$.

After this path reparametrization and making a saddle point
approximation, this path integral \\ $ \tilde{J}_r(x_{1m}, x_{2m},
y_{1m}, y_{2m}, t | x_{10}, x_{20}, y_{10}, y_{20}, 0 ; [\vec{b}]
) $ can be written  as
\begin{eqnarray}
& &  \tilde{J}_r(x_{1m}, x_{2m}, y_{1m}, y_{2m}, t | x_{10}, x_{20},
y_{10}, y_{20}, 0 ; [\vec{b}] ) = \tilde{J}_r(0, 0, 0, 0, t | 0, 0,
0, 0, 0 ; [\vec{b}] ) \nonumber\\ & \times &
\exp{(\frac{i}{\hbar} A[ \Sigma_{cl,1}, \Sigma_{cl,2}, \Delta_{cl,1}, \Delta_{cl,2}
]   )} \nonumber\\
&\times & \exp{[\frac{i}{\hbar}(  - dt \int_{0}^{t} d s
J_{\vec{\Sigma}}(s) ( \Sigma_{cl,1}(s) + \Sigma_{cl,2}(s) ) + i dt
\int_{0}^{t} d s J_{\vec{\Delta}}(s) ( \Delta_{cl,1}(s) +
\Delta_{cl,2}(s)) )]},
\end{eqnarray}
where
\begin{eqnarray}
& & \tilde{J}_r(0, 0, 0, 0, t | 0, 0, 0, 0, 0 ; [\vec{b}] )
\int_{0;\varphi_1=0}^{t;\varphi_1=0} {\cal D}\varphi_{1}
\int_{0;\varphi_2=0}^{t;\varphi_2=0} {\cal D}\varphi_{2}
\int_{0;\psi_1=0}^{t;\psi_1=0} {\cal D}\psi_{1}
\int_{0;\psi_2=0}^{t;\psi_2=0} {\cal D}\psi_{2}\nonumber \\
& & \exp{\{ i[ \int_{0}^{t} d s_1 \int_{0}^{t}
\frac{1}{2}\Psi^T(s_1) \hat{O}(s_1,s_2) \Psi(s_2) + \int_{0}^{t} d s
\Psi^T(s) \cdot \vec{B}(s)  ] \}}.
\end{eqnarray}
Note that
\begin{eqnarray}
\Psi = \left(%
\begin{array}{c}
  \Psi_1 \\
  \Psi_2 \\
\end{array}%
\right) = \left(%
\begin{array}{c}
  \varphi_1 \\
  \psi_1 \\
  \varphi_2 \\
  \psi_2 \\
\end{array}%
\right)
\end{eqnarray}
and
\begin{eqnarray}
\vec{B} = \left(%
\begin{array}{c}
  -dt J_{\vec{\Sigma}} \\
  i dt J_{\vec{\Delta}} + i \tilde{J}_{\vec{\Delta}}  \\
  -dt J_{\vec{\Sigma}} \\
  i dt J_{\vec{\Delta}} + i \tilde{J}_{\vec{\Delta}} \\
\end{array}%
\right),
\end{eqnarray}
where $ \tilde{J}_{\vec{\Delta}} $ is a new source which appears
because the nonlocality of the influence functional. It couples
the classical paths to the $ \Psi $ paths.
\begin{eqnarray}
 \tilde{J}_{\vec{\Delta}}(s) = \int_{0}^{t} d s^{\prime}
 [ \Delta_{cl,1}(s^{\prime}) + \Delta_{cl,2}(s^{\prime})  ]
 \nu( s - s^{\prime} ).
\end{eqnarray}

The matrix operator $\hat{O}(s_1, s_2)$ is defined as follows:
\begin{eqnarray}
O_{11}(s_1,s_2) = O_{33}(s_1,s_2) = O_{13}(s_1,s_2) O_{31}(s_1,s_2) = 0,
\end{eqnarray}
\begin{eqnarray}
O_{22}(s_1,s_2) = O_{44}(s_1,s_2) = O_{24}(s_1,s_2) O_{42}(s_1,s_2) = 2 i \nu(s_1
-s_2),
\end{eqnarray}
\begin{eqnarray}
O_{14}(s_1,s_2) = O_{32}(s_1,s_2) = 2 \theta(s_2 - s_1) \eta(s_1 -
s_2),
\end{eqnarray}
\begin{eqnarray}
O_{41}(s_1,s_2) = O_{23}(s_1,s_2) = 2 \theta(s_1 - s_2) \eta(s_1 -
s_2),
\end{eqnarray}
\begin{eqnarray}
O_{12}(s_1,s_2) = O_{34}(s_1,s_2) = \left(\frac{d^2}{ds_1^2} +
\Omega^2 \right)\delta(s_1 - s_2) + 2 \theta(s_2 - s_1) \eta(s_1 - s_2),
\end{eqnarray}
\begin{eqnarray}
O_{21}(s_1,s_2) = O_{43}(s_1,s_2) = \left(\frac{d^2}{ds_1^2} +
\Omega^2 \right) \delta(s_1 - s_2) + 2 \theta(s_1 - s_2) \eta(s_1 -
s_2).
\end{eqnarray}

The Gaussian path integral can be computed in terms of the inverse
of the operator $\hat{O}$, which is given by $\hat{G} \equiv
\hat{O}^{-1}$.  Hence to first order in $dt$, we have
\begin{eqnarray*}
& & \tilde{J}_r(0, 0, 0, 0, t | 0, 0, 0, 0, 0 ; [\vec{b}] )
\int_{0;\varphi_1=0}^{t;\varphi_1=0} {\cal D}\varphi_{1}
\int_{0;\varphi_2=0}^{t;\varphi_2=0} {\cal D}\varphi_{2}
\int_{0;\psi_1=0}^{t;\psi_1=0} {\cal D}\psi_{1}
\int_{0;\psi_2=0}^{t;\psi_2=0} {\cal D}\psi_{2} \\
& & \exp{\{ i[ \int_{0}^{t} d s_1 \int_{0}^{t}
\frac{1}{2}\Psi^T(s_1) \hat{O}(s_1,s_2) \Psi(s_2) + \int_{0}^{t} d
s \Psi^T(s) \cdot \vec{B}(s)  ] \}} \\
&& = \int {\cal D}\varphi_{1} \int {\cal D}\varphi_{2} \int {\cal
D}\psi_{1} \int {\cal D}\psi_{2} \exp{ \{ i[\frac{1}{2}(\Psi^T +
\vec{B}^T\cdot \hat{O}^{-1})\hat{O}(\Psi + \hat{O}^{-1} \cdot
\vec{B}) -\frac{1}{2}\vec{B}^T \hat{O}^{-1}\vec{B} ] \}} \\
& = & Z_0(t) \exp{ \{  - \frac{i}{2} \vec{B}^T \hat{O}^{-1}\vec{B}
\}} \\
& \approx & Z_0(t) ( 1  - \frac{i}{2} \vec{B}^T
\hat{O}^{-1}\vec{B} ) \\
& \approx & Z_0(t)( 1 - dt \int_{0}^{t}ds_1 \int_{0}^{t}ds_2
J_{\vec{\Sigma}}(s_1)  [ G_{12}(s_1,s_2)  + G_{14}(s_1,s_2) +
 G_{21}(s_2,s_1)  + G_{41}(s_1,s_2)
 ] \tilde{J}_{\vec{\Delta}}(s_2)   ).
\end{eqnarray*}
Note that the Green`s function $(G_{12} + G_{32}) \equiv
\tilde{G}_{12}(s_1, s_2)$ satisfies the following equation
\begin{eqnarray}
\frac{d^2}{ds_1^2}\tilde{G}_{12}(s_1, s_2) + \Omega^2
\tilde{G}_{12}(s_1, s_2) + 4 \int_{0}^{s_1} d \tau \eta(s_1 - \tau)
\tilde{G}_{12}(s_1, \tau) = \delta(s_1 - s_2)
\end{eqnarray}
with boundary conditions $\tilde{G}_{12}(0, s_2)=
\tilde{G}_{12}(s_1, t) = 0$. The equation for $(G_{21} + G_{23})
\equiv \tilde{G}_{21}(s_1, s_2)$ are analogous.

Now we can show that
\begin{eqnarray}
& &  \tilde{J}_r(x_{1m}, x_{2m}, y_{1m}, y_{2m}, t | x_{10},
x_{20}, y_{10}, y_{20}, 0 ; [\vec{b}] ) \nonumber \\
& & = \tilde{J}_r(0, 0, 0, 0, t | 0, 0, 0, 0, 0 ; [\vec{b}] ) \exp{\{ i (
A[\Sigma_{cl,1}, \Sigma_{cl,2}, \Delta_{cl,1}, \Delta_{cl,2}] )\} }\nonumber \\
&&\times \exp {\{i( - dt \int_{0}^{t} d s J_{\vec{\Sigma}}(s) (
\Sigma_{cl,1}(s) + \Sigma_{cl,2}(s)  ) + i dt \int_{0}^{t} d s
J_{\vec{\Delta}}(s) ( \Delta_{cl,1}(s) + \Delta_{cl,2}(s)  ) ) \}
 } \\
 &&\approx Z_0(t) \exp{ \{ i A[\Sigma_{cl,1}, \Sigma_{cl,2}, \Delta_{cl,1},
\Delta_{cl,2}]  \}  } \\
&&\times  \{ 1 -  dt \int_{0}^{t}ds_1 \int_{0}^{t}ds_2
J_{\vec{\Sigma}}(s_1)  [ \tilde{G}_{12}(s_1,s_2)  +
 \tilde{G}_{21}(s_1,s_2)
 ] \tilde{J}_{\vec{\Delta}}(s_2)  \\ && - i dt \int_{0}^{t} d s J_{\vec{\Sigma}}(s) (
\Sigma_{cl,1}(s) + \Sigma_{cl,2}(s)  ) + (i)^2 dt \int_{0}^{t} d s
J_{\vec{\Delta}}(s) ( \Delta_{cl,1}(s) + \Delta_{cl,2}(s)  ) \} \\
&& = J_r(x_{1m}, x_{2m}, y_{1m}, y_{2m}, t | x_{10}, x_{20},
y_{10}, y_{20}, 0 )\nonumber \\
&& \times W(x_{1m}, x_{2m}, y_{1m}, y_{2m}, x_{10}, x_{20}, y_{10},
y_{20}, dt ),
\end{eqnarray}
where $W$ is given by,
\begin{eqnarray}
&& W(x_{1m}, x_{2m}, y_{1m}, y_{2m}, x_{10}, x_{20}, y_{10},
y_{20}, dt )  \nonumber \\
&=& 1 - i dt [ \int_{0}^{t} d s 2 (\Delta_{1f} + \Delta_{2f}) \eta
(t - s)u_1(s) \tilde{\Sigma}_{cl}(0) \nonumber\\
&&+ \int_{0}^{t} d s 2 (\Delta_{1f} + \Delta_{2f}) \eta (t -
s)u_2(s)
\tilde{\Sigma}_{cl}(t)  ] \\
&-& dt [ \int_{0}^{t} d s  (\Delta_{1f} + \Delta_{2f}) \nu (t - s)
v_1(s) \tilde{\Delta}_{cl}(0) + \int_{0}^{t} d s  (\Delta_{1f} +
\Delta_{2f}) \nu (t - s)v_2(s) \tilde{\Delta}_{cl}(t)   ] \\
&-& dt [ \int_{0}^{t}ds_1 \int_{0}^{t}ds_2 \int_{0}^{t}ds_3 2
(\Delta_{1f} + \Delta_{2f}) \eta (t - s_1)  [
\tilde{G}_{12}(s_1,s_2) +
 \tilde{G}_{21}(s_2,s_1)
 ] \nonumber \\
&&\times \nu(s_2 - s_3) v_1(s_3) \tilde{\Delta}_{cl}(0) \\
 &+&  \int_{0}^{t}ds_1 \int_{0}^{t}ds_2 \int_{0}^{t}ds_3 2
(\Delta_{1f} + \Delta_{2f}) \eta (t - s_1)  [
\tilde{G}_{12}(s_1,s_2) \nonumber \\
&& +
 \tilde{G}_{21}(s_2,s_1)
 ] \nu(s_2 - s_3) v_2(s_3) \tilde{\Delta}_{cl}(t)     ].
\end{eqnarray}

To simplify the expressions, let us define
\begin{eqnarray}
d_1(t) &=& 2 \int_{0}^{t} d s \eta(t - s) u_1(s), \quad d_2(t) = 2
\int_{0}^{t} d s \eta(t - s) u_2(s),\\
 c_1(t) &=& \int_{0}^{t}ds_1
\int_{0}^{t}ds_2 \int_{0}^{t}ds_3 \eta (t - s_1)  [
\tilde{G}_{12}(s_1,s_2) +
 \tilde{G}_{21}(s_2,s_1)
 ] \nu(s_2 - s_3) v_1(s_3) , \\
 c_2(t) &=& \int_{0}^{t}ds_1 \int_{0}^{t}ds_2 \int_{0}^{t}ds_3
\eta (t - s_1)  [ \tilde{G}_{12}(s_1,s_2) +
 \tilde{G}_{21}(s_2,s_1)
 ] \nu(s_2 - s_3) v_2(s_3) ,\\
e_1(t) &=&  \int_{0}^{t} d s \nu(t - s) v_2(s) = \int_{0}^{t} d s
\nu(t - s) u_1(t -s) = \int_{0}^{t} d s \nu( s) u_1(s) \\
e_2(t) &=&  \int_{0}^{t} d s \nu(t - s) v_1(s) = \int_{0}^{t} d s
\nu(t - s) u_2(t -s) = \int_{0}^{t} d s \nu( s) u_2(s).
\end{eqnarray}

Finally, we have,
\begin{eqnarray}
& & J_r(x_{1f}, x_{2f}, y_{1f}, y_{2f}, t+dt | x_{10}, x_{20},
y_{10}, y_{20}, 0 ) \nonumber \\
 & = &  N(t)
\prod^{2}_{k=1}\int_{-\infty}^{\infty} d \beta_{kx}
   \int_{-\infty}^{\infty} d \beta_{ky} \exp{ ( \frac{i }{2  dt} (\beta_{1x}^2 +
\beta_{2x}^2 -\beta_{1y}^2
-\beta_{2y}^2 ) )  } \nonumber  \\
& \times & \{ 1 -  dt  [ i(V(x_{1f}, x_{2f}) - V(y_{1f}, y_{2f}))
 + i  (\Delta_{1f} + \Delta_{2f} )  ( d_1(t)
(\Sigma_{i,1} + \Sigma_{i,2} ) \nonumber \\
&&+ d_2(t)(\Sigma_{1f} + \Sigma_{2f}
))   + (\Delta_{1f} + \Delta_{2f} )(\Delta_{i,1} +
\Delta_{i,2} )( e_2(t) + 2 c_1(t) ) \nonumber \\
&&+ (\Delta_{1f} + \Delta_{2f}
)^2 ( e_1(t) + 2 c_2(t) )      ] \} \nonumber \\
&& \times \{  J_r + \frac{1}{2}[    \frac{\partial^2 J_r}{\partial
x_{1f}^2}(-\beta_{1x})^2  + \frac{\partial^2 J_r}{\partial
x_{2f}^2}(-\beta_{2x})^2  + \frac{\partial^2 J_r}{\partial
y_{1f}^2}(-\beta_{1y})^2  + \frac{\partial^2 J_r}{\partial
y_{2f}^2}(-\beta_{2y})^2         ]       \}.
\end{eqnarray}
Hence
\begin{eqnarray}
&& J_r(t+dt) -J_r= -dt J_r[ i \frac{1}{2}\Omega^2( x_{1f}^2 +
x_{2f}^2 -  y_{1f}^2 -  y_{2f}^2 ) + (\Delta_{1f} + \Delta_{2f} )
\times \nonumber  \\ && [ i ( d_1(t) (\Sigma_{i,1} + \Sigma_{i,2} ) +
d_2(t)(\Sigma_{1f} + \Sigma_{2f} ))   + (\Delta_{i,1} +
\Delta_{i,2} )( e_2(t) + 2 c_1(t) )\nonumber \\
&& + (\Delta_{1f} + \Delta_{2f} )
( e_1(t) + 2 c_2(t) )    ]] \\
&& + \frac{1}{2} \frac{dt}{-i} \frac{\partial^2J_r}{\partial
x_{1f}^2} + \frac{1}{2} \frac{dt}{-i}
\frac{\partial^2J_r}{\partial x_{2f}^2} + \frac{1}{2} \frac{dt}{i}
\frac{\partial^2J_r}{\partial y_{1f}^2} + \frac{1}{2} \frac{dt}{i}
\frac{\partial^2J_r}{\partial y_{2f}^2},
\end{eqnarray}
We can then get the evolution equation for the propagator $J_r$.
\begin{eqnarray}
&& i \frac{\partial}{\partial t} J_r(x_{1f}, x_{2f}, y_{1f},
y_{2f}, t | x_{10}, x_{20}, y_{10}, y_{20}, 0 )\nonumber \\
&=& i \frac{\partial}{\partial t} [ J_r(x_{1f}, x_{2f}, y_{1f},
y_{2f}, t+dt | x_{10}, x_{20}, y_{10}, y_{20}, 0 ) - J_r(x_{1f},
x_{2f}, y_{1f}, y_{2f}, t | x_{10}, x_{20}, y_{10}, y_{20}, 0 )
]\nonumber\\
& = & \{ - \frac{1}{2}( \frac{\partial^2}{\partial x_{1f}^2} +
\frac{\partial^2}{\partial x_{2f}^2} - \frac{\partial^2}{\partial
y_{1f}^2} - \frac{\partial^2}{\partial y_{2f}^2} ) +
\frac{1}{2}\Omega^2( x_{1f}^2 + x_{2f}^2 -  y_{1f}^2 -  y_{2f}^2 )
\nonumber\\ &+&  (\Delta_{1f} + \Delta_{2f} )( ( d_1(t)
(\Sigma_{i,1} +
\Sigma_{i,2} ) + d_2(t)(\Sigma_{1f} + \Sigma_{2f} ))  ) \nonumber\\
&-& i (\Delta_{1f} + \Delta_{2f} ) (\Delta_{i,1} + \Delta_{i,2} )(
e_2(t) + 2 c_1(t) ) \nonumber\\ &-& i (\Delta_{1f} + \Delta_{2f}
)^2( e_1(t) + 2 c_2(t) )
        \}J_r(x_{1f}, x_{2f}, y_{1f}, y_{2f}, t | x_{10}, x_{20}, y_{10},
y_{20}, 0 ).
\end{eqnarray}

\subsection{ Step four} Now we have the explicit expression for
$J_r$. But we still need to deal with  terms of the form like
$\Delta_{1i} J$. To do so we can differentiate $J$ with respect to
$\Sigma_{1f}$ and get
\begin{eqnarray}
\partial_{\Sigma_{1f}} J &=& [ib_1(t) (\Delta_{1f} + \Delta_{2f}) + ib_5(t)
(\Delta_{1f} \nonumber\\
&&- \Delta_{2f}) - ib_3(t) (\Delta_{1i} + \Delta_{2i}) - ib_7(t)
(\Delta_{1i} - \Delta_{2i}) ]J.
\end{eqnarray}
Similarly if we want $\Delta_{2i} J$, we can differentiate $J$ with
respect to $\Sigma_{2f}$ and get
\begin{eqnarray}
\partial_{\Sigma_{2f}} J &=& [ib_1(t) (\Delta_{1f} + \Delta_{2f}) - ib_5(t)
(\Delta_{1f}\nonumber\\
&& - \Delta_{2f}) - ib_3(t) (\Delta_{1i} + \Delta_{2i}) + ib_7(t)
(\Delta_{1i} - \Delta_{2i}) ]J.
\end{eqnarray}
The sum of these two equations gives
\begin{eqnarray}
(\partial_{\Sigma_{1f}} + \partial_{\Sigma_{2f}}) J = [2 ib_1(t)
(\Delta_{1f} + \Delta_{2f}) - 2ib_3(t) (\Delta_{1i} + \Delta_{2i})
 ]J.
\end{eqnarray}
This can be written as
\begin{eqnarray}
(\Delta_{1i} + \Delta_{2i})J = \frac{1}{2 b_3(t)}
[i(\partial_{\Sigma_{1f}} + \partial_{\Sigma_{2f}}) + 2 b_1(t)
(\Delta_{1f} + \Delta_{2f}) ]J.
\end{eqnarray}
Similarly, we can differentiate with respect to $\Delta_{1f}$ (or $
\Delta_{2f}$ )to get $\Sigma_{1i}J$ (or $\Sigma_{2i}J$). The sum of
these two equations gives
\begin{eqnarray}
( \partial_{\Delta_{1f}} + \partial_{\Delta_{2f}}) J &=& 2[ib_2(t)
(\Sigma_{1i} + \Sigma_{2i}) + ib_1(t) (\Sigma_{1f} +
\Sigma_{2f})\nonumber
\\ &-& a_{12}(t)(\Delta_{1i} + \Delta_{2i}) -2
a_{11}(t)(\Delta_{1f} + \Delta_{2f})  ]J
\end{eqnarray}
and
\begin{eqnarray}
(\Sigma_{1i} + \Sigma_{2i}) J &=& \frac{1}{2 b_2(t)}
[-i(\partial_{\Delta_{1f}} + \partial_{\Delta_{2f}}) +
\frac{a_{12}(t)}{b_3(t)} (\partial_{\Sigma_{1f}} +
\partial_{\Sigma_{2f}}) - 2 b_1(t) (\Sigma_{1f} + \Sigma_{2f})\nonumber\\
 && -i
[4a_{11}(t) + 2\frac{a_{12}(t)b_1(t)}{b_3(t)} ](\Delta_{1f} +
\Delta_{2f}) ]J.
\end{eqnarray}

Substituting in what we already have for $(\Sigma_{1i} +
\Sigma_{2i})J$ and $(\Delta_{1i} + \Delta_{2i})J$, and multiplying
by $\rho_0$ and integrating over initial coordinates, we obtain
\begin{eqnarray}
&&(\Delta_{1f} + \Delta_{2f}) d_1(t)(\Sigma_{1i} + \Sigma_{2i}) J\nonumber \\
&=& (\Delta_{1f} + \Delta_{2f}) d_1(t) [\frac{-i}{2
b_2(t)}(\partial_{\Delta_{1f}} +
\partial_{\Delta_{2f}}) + \frac{a_{12}(t)}{2 b_2(t) b_3(t)}
(\partial_{\Sigma_{1f}} +
\partial_{\Sigma_{2f}}) \nonumber \\
&-&  \frac{b_1(t)}{b_2(t)} (\Sigma_{1f} + \Sigma_{2f})
 -i
[\frac{2a_{11}(t)}{b_2(t)} + \frac{a_{12}(t)b_1(t)}{b_2(t) b_3(t)}
](\Delta_{1f} + \Delta_{2f}) ]J,
\end{eqnarray}
and
\begin{eqnarray}
&& (\Delta_{1f} + \Delta_{2f})( e_2(t) +2 c_1(t) ) (\Delta_{1i} +
\Delta_{2i}) J\nonumber \\
&=& (\Delta_{1f} + \Delta_{2f}) ( e_2(t) +2 c_1(t) ) [\frac{i}{2
b_3(t)}(\partial_{\Sigma_{1f}} + \partial_{\Sigma_{2f}}) +
\frac{b_1(t)}{ b_3(t)} (\Delta_{1f} + \Delta_{2f}) ]J
\end{eqnarray}
Hence we can write the evolution equation for the reduced density
matrix as
\begin{eqnarray}
i \frac{\partial}{\partial t} \rho_r &=& [
-\frac{1}{2}(\frac{\partial^2}{\partial x_1^2} +
\frac{\partial^2}{\partial x_2^2} - \frac{\partial^2}{\partial
y_1^2} - \frac{\partial^2}{\partial y_2^2} ) + \frac{1}{2}\Omega^2
(x_1^2 + x_2^2 - y_1^2 - y_2^2 ) ]\rho_r \nonumber \\
&& + \delta \Omega^2(t)(\Delta_{1f} + \Delta_{2f})(\Sigma_{1f} +
\Sigma_{2f})\rho_r \nonumber\\
&& -i A_1(t)(\Delta_{1f} + \Delta_{2f})(\partial_{\Delta_{1f}} +
\partial_{\Delta_{2f}})\rho_r \nonumber\\
&& -i A_2(t)(\Delta_{1f} + \Delta_{2f})^2 \rho_r\nonumber \\
&& + A_3(t)(\Delta_{1f} + \Delta_{2f})(\partial_{\Sigma_{1f}} +
\partial_{\Sigma_{2f}})\rho_r
\end{eqnarray}
where
\begin{eqnarray}
\frac{\partial}{\partial \Sigma} = \frac{\partial}{\partial x} +
\frac{\partial}{\partial y}; \quad \frac{\partial}{\partial
\Delta} = \frac{1}{2} (\frac{\partial}{\partial x} -
\frac{\partial}{\partial y})
\end{eqnarray}
and
\begin{eqnarray}
\delta \Omega^2(t) \equiv d_2(t) -d_1(t)\frac{b_1(t)}{b_2(t)},
\quad A_1(t) \equiv \frac{d_1(t)}{2b_2(t)},
\end{eqnarray}
\begin{eqnarray}
A_2(t) \equiv d_1(t)[\frac{2a_{11}(t)}{b_2(t)} + \frac{a_{12}(t)
b_1(t)}{b_2(t) b_3(t)}] + (e_1(t) + 2c_2(t)) + (e_2(t) +
2c_1(t))\frac{b_1(t)}{b_3(t)}
\end{eqnarray}
\begin{eqnarray}
A_3(t) \equiv \frac{d_1(t) a_{12}(t) }{2 b_2(t) b_3(t)} +
\frac{e_2(t) + 2c_1(t)}{2b_3(t)}.
\end{eqnarray}
This immediately leads to the general master equation
(\ref{mastercoord}).

\subsection{Coefficients of the Master Equation}

The determination of the coefficients is reasonably standard, so we
only provide the explicit forms of those time-dependent functions
that will be used later on.  As shown in \cite{HPZ}, the functions
$\delta\Omega^2(t), \Gamma(t), \Delta(t), \Sigma(t)$ can be
constructed in terms of the elementary functions $u_i(s), i=1,2$,
which satisfy the following homogeneous integro-differential
equation:
\begin{equation}{\ddot f}(s)+\Omega^2 f(s)+{4\over M}\int_0
^s\,d\lambda\eta(s-\lambda)f(\lambda)=0
\end{equation}
with  the boundary conditions:
\begin{equation}u_1(s=0)=1\>,\,\,  u_1(s=t)=0\> ,
\end{equation}
and
\begin{equation}u_2(s=0)=0\>,\,\,  u_2(s=t)=1.
\end{equation}
Here  $\eta(t-s)$ is the dissipation kernel given by
\begin{equation}
\eta(s)= -\int_0^{\infty} d\omega I(\omega) \sin(\omega s),
\end{equation}
and $I(\omega)$ is the spectral density of the environment. Note that
the numerical factor 4  before the integral in this equation is
different from that in \cite{HPZ}. This is the main difference due to
the presence of two harmonic oscillators. Although the two harmonic
oscillators are not coupled directly, they are connected by the
common reservoir, hence they affect each other dynamically.

Let $G_1(s,\tau)$ be the Green function obeying the following
equation:
\begin{equation}
\frac{d^2}{ds^2}G_1(s,\tau) + \Omega^2 G_1(s,\tau) +
\frac{4}{M}\int_0^s d \tau \eta(s-\tau)G_1(s,\tau) =\delta(s-\tau),
\end{equation}
with initial conditions:
\begin{equation}
G_1(s=0,\tau)= 0\>,\> \> {d\over ds}G_1(s,\tau)|_{s=0}=0\>.
\end{equation}
The Green function $G_2(s, \tau)$ is defined analogously. The
coefficients can then be written as
\begin{eqnarray}
\delta \Omega^2(t) &=&
\frac{2}{M}\int^t_0ds\eta(t-s)\left(u_2(s)-{u_1(s) \dot u_2(t)\over
\dot
 u_1(t)}\right),\\
 \Gamma(t) &=& {1 \over M}\int^t_0ds\eta(t-s)\frac{u_1(s)}{\dot u_1(t)},
\end{eqnarray}
\begin{eqnarray}
\Delta(t)&=&{\hbar\over 2M}\int^t_0\,d\lambda
G_1(t,\lambda)\nu(t-\lambda)\nonumber
\\
    & & -{4\hbar\over M^2}\int^t_0ds\int^t_sd\tau \int^t_0d\lambda
       \eta(t-s)G_1(t,\lambda)G_2(s,\tau)\nu(\tau-\lambda),
\end{eqnarray}
and
\begin{eqnarray}
\Sigma(t)&=&\frac{\hbar}{2}\int^t_0\,d\lambda G'_1(t,\lambda)\nu(t-\lambda)\nonumber \\
    & & -{4\hbar\over M}\int^t_0ds\int^t_sd\tau \int^t_0d\lambda
        \eta(t-s)G'_1(t,\lambda)G_2(s,\tau)\nu(\tau-\lambda),
\end{eqnarray}
where $\nu(s)$  defined as
\begin{equation}
\nu(s)=\int^{+\infty}_0d\omega I(\omega)\coth({1\over
2}\hbar\omega\beta)\cos(\omega s),
\end{equation}
 is the noise kernel of the environment. Here a ``prime'' denotes
taking the derivative with respect to the first variable of
$G_1(s,\tau)$.

\section{Explicit Expressions For  $\rho_{ij}$}

We find that the matrix $G_{ij}$ is the same for all the
$\rho_{ij}$. Thus, we can write $G_{ij}  \equiv  G$.  The matrix elements for the
matrix $G$ are given by
 \beqa
 G_{11}&=&G_{22} = a_{22} + \frac{i b_4}{2} + \frac{i b_8}{2} + \frac{1}{2 \delta^2},\\
 G_{33}&=&G_{44} = a_{22} - \frac{i b_4}{2} - \frac{i b_8}{2} + \frac{1}{2 \delta^2},\\
 G_{12}&=&G_{21}= \frac{1}{2}(2 a_{22} + i b_4 - i b_8),\\
 G_{34}&=&G_{43}= \frac{1}{2}(2 a_{22} - i b_4 + i b_8),  \\
G_{13}&=& G_{14}= G_{23}=G_{24} = G_{31}=G_{32}= G_{41}=G_{42} = - a_{22}.
\eeqa
Then the determinant of $G$ can be explicitly computed,
\begin{eqnarray}
\det G = b_4^2 b_8^2 + \frac{1}{16 \delta^8} + \frac{a_{22}}{2
\delta^6} + \frac{b_{4}^2}{2 \delta^4} + \frac{b_{8}^2}{4 \delta^4}
+ \frac{2 a_{22} b_8^2}{ \delta^2}.
\end{eqnarray}
Moreover, the matrix elements of the inverse matrix $G^{-1}$ are
\begin{eqnarray}
G^{-1}_{11}=G^{-1}_{22} &=& \frac{1}{\det G}(-\frac{i}{2}b_4^2 b_8 +
a_{22}b_8^2 \nonumber \\
&&- \frac{i}{2}b_4 b_8^2 + \frac{1}{8 \delta^6} +
\frac{3 a_{22}}{4\delta^4} - \frac{i b_4}{8\delta^4} - \frac{i
b_8}{8\delta^4} +  \frac{ b_4^2}{4 \delta^2} - \frac{i a_{22}
b_8}{\delta^2} +  \frac{ b_8^2}{4 \delta^2} ),\\
G^{-1}_{33}=G^{-1}_{44}&=& \frac{1}{\det G}(\frac{i}{2}b_4^2 b_8 +
a_{22}b_8^2  \nonumber \\
&&+ \frac{i}{2}b_4 b_8^2 + \frac{1}{8 \delta^6} +
\frac{3 a_{22}}{4\delta^4} + \frac{i b_4}{8\delta^4} + \frac{i
b_8}{8\delta^4} +  \frac{ b_4^2}{4 \delta^2} + \frac{i a_{22}
b_8}{\delta^2} +  \frac{ b_8^2}{4 \delta^2} ),\\
G^{-1}_{12}=G^{-1}_{21}&=& \frac{1}{\det G}(\frac{i}{2}b_4^2 b_8 +
a_{22}b_8^2 \ \nonumber \\
&&- \frac{i}{2}b_4 b_8^2 - \frac{a_{22}}{4 \delta^4} -
\frac{i b_4}{8\delta^4} + \frac{i b_8}{8\delta^4} - \frac{
b_4^2}{4 \delta^2} + \frac{i a_{22} b_8}{\delta^2} + \frac{
b_8^2}{4 \delta^2} ),\\
G^{-1}_{34}=G^{-1}_{43}&=& \frac{1}{\det G}(- \frac{i}{2}b_4^2 b_8 +
a_{22}b_8^2  \nonumber \\
&&+ \frac{i}{2}b_4 b_8^2 - \frac{a_{22}}{4 \delta^4} +
\frac{i b_4}{8\delta^4} - \frac{i b_8}{8\delta^4} - \frac{
b_4^2}{4 \delta^2} - \frac{i a_{22} b_8}{\delta^2} + \frac{
b_8^2}{4 \delta^2} ),\\
G^{-1}_{13}=G^{-1}_{14}G^{-1}_{23}&=&G^{-1}_{24}=G^{-1}_{31}=G^{-1}_{32}=G^{-1}_{41}=G^{-1}_{42}=\frac{1}{\det
G}( a_{22}b_8^2 + \frac{a_{22}}{4 \delta^4}  ).
\end{eqnarray}

For the case of $\rho_{11}$:
\begin{eqnarray}
\rho_{11}(t=0) &=& N^4 \exp{[- \frac{(x_{10} - L_0)^2 + (x_{20} -
L_0)^2 + (y_{10} - L_0)^2 + (y_{20} - L_0)^2}{2 \delta^2}]} \nonumber \\
&\times& \exp {[ i P_0( x_{10} + x_{20} - y_{10} - y_{20}) ] },
\end{eqnarray}
then the matrix elements for $F$ are,
\begin{eqnarray*}
F^1_{11}&=& i P_0 -a_{12} x_1 +\frac{i b_2 x_1}{2} - \frac{i b_3
x_1}{2}+ \frac{i b_6 x_1}{2} - \frac{i b_7 x_1}{2} - a_{12} x_2
+\frac{i b_2 x_2}{2} - \frac{i b_3 x_2}{2} - \frac{i b_6 x_2}{2} +
\frac{i b_7 x_2}{2}\nonumber\\ &+&  a_{12} y_1 - \frac{i b_2 y_1}{2}
- \frac{i b_3 y_1}{2} - \frac{i b_6 y_1}{2} - \frac{i b_7 y_1}{2} +
a_{12} y_2 - \frac{i b_2 y_2}{2} - \frac{i b_3 y_2}{2} + \frac{i b_6
y_2}{2} + \frac{i b_7 y_2}{2} + \frac{L_0}{\delta^2}\\
F^2_{11}= i P_0 &-& a_{12} x_1 +\frac{i b_2 x_1}{2} - \frac{i b_3
x_1}{2} - \frac{i b_6 x_1}{2} + \frac{i b_7 x_1}{2} -
a_{12}x_2+\frac{i b_2 x_2}{2}
- \frac{i b_3 x_2}{2} + \frac{i b_6 x_2}{2} - \frac{i b_7 x_2}{2}\\
&+&  a_{12} y_1 - \frac{i b_2 y_1}{2} - \frac{i b_3 y_1}{2} +
\frac{i b_6 y_1}{2} + \frac{i b_7 y_1}{2} + a_{12} y_2 - \frac{i b_2
y_2}{2} - \frac{i b_3 y_2}{2} - \frac{i b_6 y_2}{2} - \frac{i b_7
y_2}{2} + \frac{L_0}{\delta^2}\\
 F^3_{11}= - i P_0 &+& a_{12} x_1
+\frac{i b_2 x_1}{2} + \frac{i b_3 x_1}{2} + \frac{i b_6 x_1}{2} +
\frac{i b_7 x_1}{2} + a_{12} x_2 +\frac{i b_2 x_2}{2} + \frac{i b_3
x_2}{2} - \frac{i b_6 x_2}{2} - \frac{i b_7 x_2}{2}\\ &-&  a_{12}
y_1 - \frac{i b_2 y_1}{2} + \frac{i b_3 y_1}{2} - \frac{i b_6
y_1}{2} + \frac{i b_7 y_1}{2} - a_{12} y_2 - \frac{i b_2 y_2}{2} +
\frac{i b_3 y_2}{2} + \frac{i b_6 y_2}{2} - \frac{i b_7 y_2}{2} +
\frac{L_0}{\delta^2}\\
F^4_{11}= - i P_0 &+& a_{12} x_1 +\frac{i b_2 x_1}{2} + \frac{i
b_3 x_1}{2} - \frac{i b_6 x_1}{2} - \frac{i b_7 x_1}{2} + a_{12}
x_2 +\frac{i b_2 x_2}{2} + \frac{i b_3 x_2}{2} + \frac{i b_6
x_2}{2} + \frac{i b_7 x_2}{2}\\ &-&  a_{12} y_1 - \frac{i b_2
y_1}{2} + \frac{i b_3 y_1}{2} + \frac{i b_6 y_1}{2} - \frac{i b_7
y_1}{2} - a_{12} y_2 - \frac{i b_2 y_2}{2} + \frac{i b_3 y_2}{2} -
\frac{i b_6 y_2}{2} + \frac{i b_7 y_2}{2} + \frac{L_0}{\delta^2} ,
\end{eqnarray*}
where $F^T_{11}= (F^1_{11}, F^2_{11}, F^3_{11}, F^4_{11})$ and
\begin{eqnarray}
c_{11}&=& - a_{11}x_1^2 + \frac{i}{2} b_1 x_1^2 + \frac{i}{2} b_5
x_1^2 - 2 a_{11} x_1 x_2 + i b_1 x_1 x_2 - i b_5 x_1 x_2 \nonumber \\
\nonumber &-& a_{11}x_2^2 + \frac{i}{2} b_1 x_2^2 + \frac{i}{2}
b_5 x_2^2 + 2 a_{11} x_1 y_1 + 2 a_{11} x_2 y_1 \\ \nonumber &-&
a_{11}y_1^2 - \frac{i}{2} b_1 y_1^2 - \frac{i}{2} b_5 y_1^2 + 2
a_{11} x_1 y_2 + i b_1 x_1 y_2 +2 a_{11} x_2 y_2 +i b_1 x_2 y_2 \\
 &-& 2 a_{11} y_1 y_2 + i b_5 y_1 y_2 - a_{11}y_2^2 +
\frac{i}{2} b_1 y_2^2 - \frac{i}{2} b_5 y_2^2 - \frac{2
L_0^2}{\delta^2} .
\end{eqnarray}

For the case of $\rho_{12}$:
\begin{eqnarray}
\rho_{12}(t=0) &=& N^4 \exp{[- \frac{(x_{10} - L_0)^2 + (x_{20} -
L_0)^2 + (y_{10} - L_0)^2 + (y_{20} + L_0)^2}{2 \delta^2}]} \nonumber \\
&\times& \exp {[ i P_0( x_{10} + x_{20} - y_{10} + y_{20}) ] }
\end{eqnarray}
\begin{eqnarray}
F^1_{12}=F^1_{11}, \quad F^2_{12}=F^2_{11},  \quad
F^3_{12}=F^3_{11},  \quad F^4_{12}=F^4_{11}+ 2i P_0 - 2
\frac{L_0}{\delta^2}, \quad  c_{12}=c_{11}.
\end{eqnarray}

For the case of $\rho_{13}$:
\begin{eqnarray}
\rho_{13}(t=0) &=& N^4 \exp{[- \frac{(x_{10} - L_0)^2 + (x_{20} -
L_0)^2 + (y_{10} + L_0)^2 + (y_{20} - L_0)^2}{2 \delta^2}]} \nonumber \\
&\times& \exp {[ i P_0( x_{10} + x_{20} + y_{10} - y_{20}) ] }
\end{eqnarray}
\begin{eqnarray}
F^1_{13}=F^1_{11}, \quad F^2_{13}=F^2_{11},  \quad
F^3_{13}=F^3_{11}+ 2i P_0 - 2 \frac{L_0}{\delta^2},  \quad
F^4_{13}=F^4_{11}, \quad  c_{13}=c_{11}.
\end{eqnarray}

For the case of $\rho_{14}$:
\begin{eqnarray}
\rho_{14}(t=0) &=& N^4 \exp{[- \frac{(x_{10} - L_0)^2 + (x_{20} -
L_0)^2 + (y_{10} + L_0)^2 + (y_{20} + L_0)^2}{2 \delta^2}]} \nonumber \\
&\times& \exp {[ i P_0( x_{10} + x_{20} + y_{10} + y_{20}) ] }
\end{eqnarray}
\begin{eqnarray}
F^1_{14}=F^1_{11}, \quad F^2_{14}=F^2_{11},  \quad
F^3_{14}=F^3_{11}+ 2i P_0 - 2 \frac{L_0}{\delta^2},  \quad
F^4_{14}=F^4_{11}+ 2i P_0 - 2 \frac{L_0}{\delta^2}, \quad
c_{14}=c_{11}.
\end{eqnarray}

Similarly, one can work out the cases for $\rho_{2i}$, $\rho_{3i}$ and  $\rho_{4i}
\,\, (i=1,2,3,4)$.
\end{appendix}


\begin{thebibliography}{99}

\bibitem{Arndt} M. Arndt {\em et. al.}, Nature {\bf 401}, 680 (1999).

\bibitem{BJK}
S. Bose, K. Jacobs and P. L. Knight, Phys. Rev.  A {\bf 59}, 3204
(1999).

\bibitem{Friedman}
J. Friedman {\em et. al.} Nature {\bf 406} 43 (2000); C. H. van
der Wall {\em et. al} Science {\bf 290} 773 (2000).

\bibitem{Armour}
A. D. Armour, M. P. Blencowe and K. C. Schwab, Phys. Rev. lett.
{\bf 88}, 148301 (2002).

\bibitem{Eisert} J. Eisert, M. B. Plenio, S. Bose,  and J. Hartley, Phys. Rev. Lett.
{\bf 93}, 190402 (2004).

\bibitem{Mancini02}
S. Mancini, V. Giovannetti, D. Vitali, and P. Tombesi, Phys. Rev.
Lett. {\bf 88}, 120401 (2002).

\bibitem{Mancini03}
S. Mancini,  D. Vitali, and P. Tombesi, Phys. Rev. Lett. {\bf 90},
137901 (2003).

\bibitem{Marshall}
W. Marshall, C. Simon, R. Penrose and D. Bouwmeester, Phys. Rev.
Lett. {\bf 91}, 130401 (2003).

\bibitem{Adler} A.Bassi, E. Ippoliti and S. L. Adler, Phys. Rev. Lett. {\bf
94} 030401 (2005).

\bibitem{Pinard} M. Pinard et al, Europhys. Lett. {\bf 72} (5) 747 (2005).

\bibitem{KRCSV} T. J. Kippenberg, H. Rokhsari, T. Carmon, A. Scherer, and K. J.
Vahala, Phys. Rev. Lett. {\bf 95}, 033901
(2005).

\bibitem{FGV} A. Ferreira, A. Guerreiro, and V. Vedral, Phys. Rev. Lett.
{\bf 96},  060407(2006).


\bibitem{Bose} S. Bose,  Phys. Rev. Lett.  {\bf 96},  060402 (2006).

 \bibitem{Blencowe} E.  Buks and M. P. Blencowe,  quant-ph/0607106.

\bibitem{Burnettetaal2007} D.  W. Hallwood, K.  Burnett, and J.  Dunningham,
quant-ph/0609077; J.A. Dunningham, K. Burnett, R. Roth, and W.D.
Phillips, quant-ph/0608242.


\bibitem {QBM}R.  Feynman and F. L. Vernon,  Ann. Phys.  (N.Y.) {\bf 24}, 118 (1963).


\bibitem{CalLeg}  A. O. Caldeira and A. J. Leggett, Physica A \textbf{121}, 587
(1983).

\bibitem {QBM1} A sample of references is listed here:
 V. Hakim and V. Ambegaokar, Phys. Rev. A \textbf{32},
423 (1985); F. Haake and R. Reibold, Phys. Rev. A \textbf{32}, 2462
(1985); W. G. Unruh and W. H. Zurek, Phys. Rev. D {\bf 40}, 1071
(1989); H. Grabert, P. Schramm, and G. L. Ingold, Phys. Rep.
\textbf{168}, 115 (1988); J.J.  Halliwell and A.  Zoupas, Phys. Rev. D {\bf 52},
7294 (1995).

\bibitem {HPZ} B. L. Hu, J. P. Paz, and Y. Zhang, Phys. Rev. D \textbf{45}, 2843
(1992); D \textbf{47}, 1576 (1993).

\bibitem{HalliwellYu96} J. J. Halliwell and T. Yu, Phys. Rev. D {\bf 53}, 2012 (1996).

\bibitem{StrunzYu2004}  W.  T.  Strunz and T. Yu, Phys. Rev.  A {\bf 69},   052115
(2004).

\bibitem{FHR}  C. H. Fleming, A. Roura and B. L. Hu, ``Solutions to Master Equations of
Quantum Brownian Motion in a General Environment with External Force"
[arXiv:0705.2766].

\bibitem{CHY07} C.H.  Chou, B. L. Hu, and T. Yu, Physica A {\bf  387}   432  (2008).

\bibitem{Kim2002} M. S.  Kim,  J.\  Lee,  D. Ahn, and P. L. Knight,  Phys. Rev. A
{\bf 65},  040101(R)  (2002).

\bibitem{Dan} D. Braun, \prl{89},  277901 (2002).

\bibitem{Kim} M. Paternostro, W. Son, and M. S. Kim,
    Phys. Rev. Lett. 92, 197901 (2004).

\bibitem{Ficek} Z. Ficek  and R. Tanas,  Phys. Rev. A {\bf 74},  024304 (2006).


\bibitem{Kubo}  R. Kubo, M. Toda, and N. Hashitsume,  {\it Statistical Physics II},
(Berlin, Springer, 1991).

\bibitem{Gardiner}C. W. Gardiner and P. Zoller, {\it Quantum Noise}
(Berlin, Springer, 2002).

\bibitem{footnote} This
 equation was originally derived for the Markovian
limit. We have added the name Markov to it since we want to call
the equation (\ref{fp} ) which covers the more general cases  the
Fokker-Planck (FP) equation. This more general equation is
sometime called the Wigner equation.

\bibitem{AH}  C. Anastopoulos
and B. L. Hu,  Phys. Rev. A {\bf 62}, 033821 (2000).

\bibitem{HZ2}  B. L. Hu and Y. Zhang,  Int. J. Mod. Phys. A {\bf 10}, 4537 (1995).

\bibitem{Yu-Eberly2004} T. Yu and J. H. Eberly,  Phys.  Rev.  Lett.   {\bf 93},
140404 (2004);  Phys. Rev. Lett. {\bf 97}, 140403 (2006).

\bibitem{Halliwell2004} P. J. Dodd  and J. J. Halliwell,  Phys. Rev. A {\bf 69},
052105 (2004).

\bibitem{PHZ} J. P. Paz, S. Habib, and W. H. Zurek, Phys. Rev. D {\bf 47}, 488 (1993).

\bibitem{duan}L. M.  Duan, G. Giedke, J. I. Cirac, and P. Zoller, Phys. Rev. Lett.
{\bf 84}, 2722 (2000).

\bibitem{simon}R.  Simon,   Phys. Rev. Lett.  {\bf 84}, 2726 (2000).

\bibitem{ASH} C. Anastopoulos, S. Shresta, and B. L. Hu, ``Quantum Entanglement
Under Non-Markovian Dynamics of Two Qubits Interacting With a Common
Electromagnetic Field",   arXiv: quant-ph/0610007.

\bibitem{LD} L. Diosi and C. Kiefer,  J. Phys.  A {\bf 35},  2675 (2002).

\bibitem{Yu-Eberly2002}  T. Yu and J. H. Eberly,  Phys.  Rev. B {\bf 66}, 193306 (2002).

\bibitem{ShiHu} K. Shiokawa and B. L. Hu, Phys. Rev. A {\bf 70}, 062106 (2004).

\bibitem{CumHu} N.  Cummings and B. L. Hu, ``Dynamics of Atom- Field Entanglement:
Towards strong coupling and non-Markovian regimes"  Phys. Rev. A
[arXiv:0708.2257].

\bibitem{LCH} S. Y. Lin, C.H. Chou and B. L. Hu, in preparation.

\bibitem{Landauer} R.  Landauer. {\em
IBM J Res Dev}, {\bf 5}, 183 (1961).


\bibitem{Bennett} C.  H. Bennett,  Studies in History and
Philosophy of Modern Physics  {\bf 34}, 501 (2003).


\bibitem{Cav93} C. M.  Caves,  {\em Phys Rev E},
{\bf 47}, 4010 (1993).

\bibitem{Bub} J. Bub,  Studies in the History and Philosophy of Modern Physics, {\bf 32},  569 ( 2001).

\bibitem{Maroney} O. J. E. Maroney, ``Generalising Landauer's Principle"
[arXiv:quant-ph/0702094]. ``The physical basis of the Gibbs-von
Neumann entropy" [arXiv:quant-ph/0701127]. ``Information and Entropy
in Quantum Theory" PhD thesis, Birkbeck College, University of
London, 2002. arXiv:quant-ph/0411172.

\bibitem{HoeBue07} C. Hoerhammer, H.  Buettner, ``Information and entropy
in quantum Brownian motion: Thermodynamic entropy versus von Neumann
entropy"  [arXiv:0710.1716].

\bibitem{HoeBue05} C. Hoerhammer, H.  Buettner, J. Phys. A {\bf 38}, 7325 (2005).

\bibitem{FleRouHu} C. H. Fleming, A. Roura and B. L. Hu,  ``Solutions to Master
Equations of Quantum Brownian Motion in a General Environment with External
Force" [arXiv:0705.2766].

\bibitem{Vitali1} D. Vitali et al, Phys. Rev. Lett.  {\bf 98}, 030405 (2007).

\bibitem{Vitali2} C. Genes, D.  Vitali, P.  Tombesi, S.
Gigan, M.  Aspelmeyer. ``Ground-state cooling of a micromechanical
oscillator: generalized framework for cold damping and
cavity-assisted cooling schemes" [Xiv:0705.1728].

\bibitem{EntSQL} H.  Mueller-Ebhardt, H.  Rehbein, R.  Schnabel, K.
Danzmann, Y.  Chen,  ``Entanglement of macroscopic test masses and
the Standard Quantum Limit in laser interferometry"
[arXiv:quant-ph/0702258].

\bibitem{YuFleHu} T. Yu,  C.  Fleming and B. L. Hu, "Master equation for Macroscopic Quantum
Phenomena in  Mirror-Photon Systems"  in preparation.

\bibitem{GalChoHu} C.  Galley, C. H. Chou and B. L. Hu, "Quantum Superposition of Two
Mirrors mediated by photons: Modeling via two harmonic oscillators in
a common bath" in preparation.


\end{thebibliography}
\end{document}